\documentclass[aps,superscriptaddress,floatfix,twocolumn,footinbib]{revtex4-1}
\usepackage{amssymb}
\usepackage{amsmath}
\usepackage{amsfonts}
\usepackage{appendix}
\usepackage{bm}
\usepackage{graphicx}
\usepackage{epsfig}
\usepackage{epstopdf}
\usepackage{balance}
\usepackage[dvipsnames]{xcolor}
\usepackage{calc}
\usepackage{natbib}
\usepackage[colorlinks,
            linkcolor=blue,
            anchorcolor=blue,
            citecolor=blue,
            urlcolor=blue]{hyperref}
\usepackage{lipsum}
\usepackage{balance}

\begin{document}

\title{ Local magnetic moment oscillation around an Anderson impurity on graphene}
\author{Shuai Li}
\affiliation{School of Physics and Wuhan National High Magnetic field center,
Huazhong University of Science and Technology, Wuhan 430074,  China}
\author{Jin-Hua Gao}
\email{jinhua@hust.edu.cn}
\affiliation{School of Physics and Wuhan National High Magnetic field center,
Huazhong University of Science and Technology, Wuhan 430074,  China}

\begin{abstract}
  We theoretically investigate the spin resolved Friedel oscillation (FO) and quasiparticle interference (QPI) in graphene induced by an Anderson impurity. Once the impurity becomes magnetic, the resulted FO becomes spin dependent, which gives rise to a local magnetic moment oscillation with an envelop decaying as $r^{-2}$ in real space in the doping cases.
  Meanwhile, at half filling, the charge density and local magnetic moment will not oscillate but decay as $r^{-3}$. Such spin resolved FO has both sublattice and spin asymmetry.
  Interestingly, the local magnetic moment decay at half filling only occurs at one sublattice of graphene, which is quite like the phenomenon observed in a recent STM experiment [H. Gonz\'alez-Herrero \textit{et al}., Science \textbf{352}, 437 (2016)]. We further give an analytic formula about such spin dependent FO based on the stationary phase approximation. Finally, we study the interference of quasiparticles around the magnetic impurity by calculating the spin-dependent Fourier-transformed local density of states (FT-LDOS). Our work  gives a comprehensive understanding about the local magnetic moment oscillation around an Anderson impurity on graphene.

\end{abstract}

\maketitle

\section{Introduction}
Friedel oscillation (FO) is the impurity induced spatial oscillation of charge density or local density of states (LDOS) in Fermi gas, \textit{e}.\textit{g}., metals, which can be understood as quasiparticle interference (QPI) pattern around the impurity~\cite{Friedel}. In experiment, FO can be directly observed with scanning tunneling microscopy (STM) by measuring the LDOS around the impurity. With the Fourier-transform (FT) STM technique,  FO provides a feasible way to  detect the dispersion relation and chirality of quasiparticles in the host metal, and has been investigated in various systems, such as metal surface states~\cite{ImageStandWave,STMSurStep,FerConFTSTM}, strongly correlated materials~\cite{Holomap,QPIStronCorr}, topological insulators~\cite{Toposurf,TRSSurf,ProBackTI,ChemDisordTI} and  2D materials~\cite{ErsiFTSTS,ChenQPI,scattinter,quasichiralprob,pseudospingraphene,foprobbystm,ftsts}.

Due to the novel massless Dirac dispersion relation~\cite{eleprograp}, the FO in graphene has been intensively studied in last decade.  Unlike the conventional 2D electron gas,  the FO in graphene has some  unusual features~\cite{scattinter,quasichiralprob,pseudospingraphene,foprobbystm,ftsts,effmonobi,ldosgraphite,FOImScaTemp,sublatticeasy,shortspapattern,FONlayer}. The key  issue is the two different impurity scattering processes, \textit{i}.\textit{e}. intervalley and intravalley scattering. At low energy, the intravalley scattering gives rise to a long wavelength FO in LDOS with envelope decaying like $r^{-2}$, which results from the suppression of backscattering due to the pseudo spin degree of freedom~\cite{effmonobi,FOImScaTemp}. The intervalley scattering is responsible for a short wavelength  oscillation of LDOS with a $r^{-1}$ decay~\cite{effmonobi,ldosgraphite}. Meanwhile, it is predicted that the FOs on the two sublattices of graphene have opposite sign, \textit{i}.\textit{e}. sublattice asymmetry~\cite{sublatticeasy}, which can essentially influence the STM  observation due to the experimental resolution.

Graphene with adsorbed atoms can be well described by the Anderson impurity model~\cite{Andimpmod}. Distinct from other kinds of defects, \textit{e}.\textit{g}., vacancy, substitutional impurity, an Anderson impurity is not a simple local potential perturbation, but can be magnetic or nonmagnetic depending on its parameters~\cite{PRL2008,zqzhangprb}. It implies that the FO resulted from an Anderson impurity should be more complex than that induced by a potential impurity, since that the spin degree of freedom has to be considered.  A typical example may be the hydrogen atom on graphene, which has been intensively studied both theoretically~\cite{hydgraelemag,undadshydgra,magstrhydgrap,ferrgrasurf,SOChydgrap,topohydgrap} and experimentally~\cite{congraprohyd,gaphydgrap,enhSOChydgrap,atomcontr}. A recent STM experiment shows that a hydrogen impurity can induce  atomically modulated spin texture in graphene around it, which extents several nanometers away from the hydrogen impurity~\cite{atomcontr}.  In our recent paper\cite{Li2019}, we have illustrated that the hydrogen adatom on graphene can be effectively viewed as an Anderson impurity, though the Coulomb interaction on carbon atoms plays an important role in making the local magentism around the adatom. To the best of our knowledge,  there is still no a detailed analysis about the Anderson impurity induced FO and quasiparticle  interference  (QPI) in graphene, despite some interesting phenomena have been observed in  experiment.

In this work, we theoretically investigate the FO and QPI induced by an Anderson impurity on graphene, and especially focus on the situations when the Anderson impurity becomes magnetic.
We first  numerically calculate the spin resolved charge density around an Anderson impurity with the self-consistent mean field method based on tight-binding (TB) model.
The numerical results indicate that, once the impurity becomes magnetic, the FO becomes spin dependent, which will give rise to  a local magnetic moment oscillation in real space and has both spin and sublattice asymmetry. An interesting case is the half filling.  At half filling, since $k_f=0$, the local magnetic moment will not oscillate but decay like $r^{-3}$ on the sublattice not directly coupled to the impurity (sublattice B),  while the local magnetic moment is tiny on the other sublattice (sublattice A).   Note that this behaviour of local magnetic moment is very like the phenomenon observed in the recent experiment~\cite{atomcontr}. With finite doping, oscillations of local magnetic moment are found on both sublattices,  and the amplitudes of the oscillations will decay as  $r^{-2}$.  We argue that this kind of local magnetic moment oscillation can be observed in experiment by fine tuning the doping.

In addition to the numerical calculation, we   give an analytic formula about this spin dependent FO on graphene in a special case, \textit{i}.\textit{e}. the  FO on sublattice A along the armchair direction. Considering the spatial symmetry of graphene, this analytic formula actually can well describe the behaviour of the spin dependent FO resulted from an Anderson impurity.  The local magnetic moment oscillation can be described by this analytic formula as well. At last, we also study  the spin-dependent QPI by calculating the Fourier-transformed LDOS (FT-LDOS). Once the impurity becomes magnetic, up and down spins feel different scattering potential, and have different FT-LDOS.

The rest of the paper is organised as follows. In Sec.~\ref{model}, we give  the relevant theoretical model and method. We discuss the numerical and analytic results of the spin dependent FO in Sec.~\ref{results}. Sec.~\ref{summary} is a brief summary.

\section{Model and method}\label{model}
\subsection{Hamiltonian}
The system we study comprises a graphene monolayer and an Anderson  impurity adsorbed on the top site of one carbon atom.
The Hamiltonian of the whole system is
\begin{equation}\label{totalH}
  H=H_0+H_{imp}+H_{hyb}.
\end{equation}
Here, $H_0$ is the Hamiltonian of graphene monolayer
\begin{equation}
  H_0=t\sum_{\langle i,j\rangle,\sigma} [ a^{\dag}_{i,\sigma}b_{j,\sigma} + H.c.],
\end{equation}
where $a_{i,\sigma}$ ($b_{j,\sigma}$) is the annihilation operator of electron with spin $\sigma=\uparrow,\downarrow$ at atomic site $i$ ($j$) of sublattice A (B), $t$ is the nearest-neighbor (NN) hopping.
The impurity is described by
\begin{equation} \label{HamilImp}
  H_{imp}=\sum_{\sigma}\varepsilon_0 d^{\dag}_{\sigma}d_{\sigma}
           +Ud^{\dag}_{\uparrow}d_{\uparrow}d^{\dag}_{\downarrow}d_{\downarrow},
\end{equation}
where $d_{\sigma}$ is the annihilation operator of impurity electron with spin $\sigma$,  $\varepsilon_0$ is the energy level of the impurity orbital, and $U$ is the on-site Coulomb interaction.
We assume that the impurity is at the origin, and coupled to an atom of sublattice A of graphene.
Then the hybridization term is
\begin{equation}
  H_{hyb}=V \sum_{\sigma} [d^{\dag}_{\sigma}a_{0,\sigma}+ H.c.],
\end{equation}
where $V$ is the coupling between the impurity and the host atom.

For the Hubbard $U$ term, we use mean-field approximation
\begin{equation}
  Ud^{\dag}_{\uparrow}d_{\uparrow}d^{\dag}_{\downarrow}d_{\downarrow} \approx U[ \langle n_{\downarrow} \rangle d^{\dag}_{\uparrow}d_{\uparrow}+ \langle n_{\uparrow} \rangle d^{\dag}_{\downarrow}d_{\downarrow}- \langle n_{\uparrow} \rangle \langle n_{\downarrow} \rangle],
\end{equation}
where  $\langle n_{\sigma} \rangle =\langle d^{\dag}_{\sigma}d_{\sigma} \rangle$.
Replacing the Hubbard term in Eq.~\eqref{HamilImp} with above approximation, the impurity level can be redefined as $\varepsilon_\sigma=\varepsilon_0+U \langle n_{\bar{\sigma}} \rangle$, which becomes spin dependent.

\subsection{$T$-matrix approach}
We use the $T$-matrix approach~\cite{effmonobi,ftsts,ldosgraphite,FONlayer,zhangzuquan2017} to calculate this impurity model. Since it is a  TB Hamiltonian in Eq.~\eqref{totalH}, the influence of  lattice structure has been considered in detail in the calculation. Note that  the $T$-matrix method here is beyond the Born approximation.

We first determine the mean fields by solving the following self-consistent equation
\begin{equation}
  \langle n_\sigma \rangle= -\frac{1}{\pi}\textmd{Im} \int^{\infty}_{-\infty}d\omega f(\omega)G_{dd,\sigma}(\omega),
\end{equation}
where $f(\omega)$ is the Fermi-Dirac distribution,
and $G_{dd,\sigma}(\omega)$ is the retarded Green's function of the impurity electrons.
$G_{dd,\sigma}(\omega)$ is given by
\begin{equation}
   G_{dd,\sigma}(\omega)=[\omega-\varepsilon_\sigma-\Sigma(\omega)+i0^{\dag}]^{-1}.
\end{equation}
Here, $\Sigma(\omega)=V^2 \textrm{g}_{00}(\omega)$ denotes the retarded self-energy, and $\textrm{g}(\omega)=(\omega-H_0+i0^{\dag})^{-1}$ is the retarded Green's function of the pristine graphene.
In real space, $\textrm{g}_{ii}(\omega)$ gives the diagonal matrix element of $\textrm{g}(\omega)$, where $i$ is the index of the atomic site.
For instance, $\textrm{g}_{00}(\omega)$ is for  the host carbon  atom at origin, to which the impurity is coupled.

\begin{figure*}
 \centering
 \includegraphics[width=17cm]{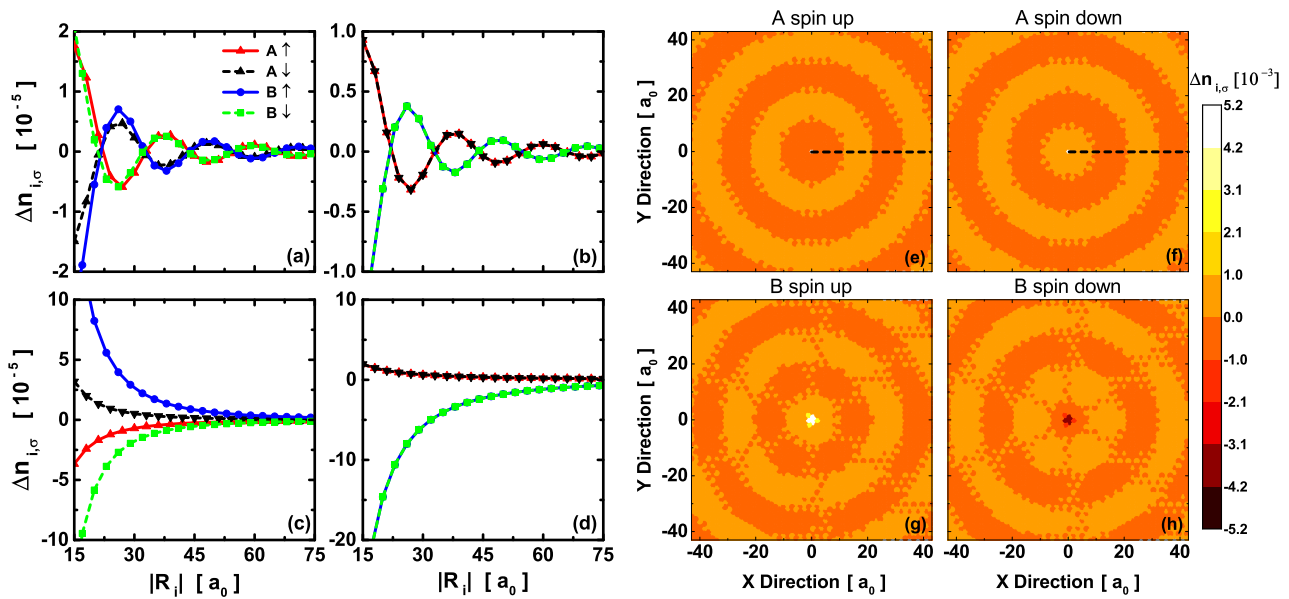}
 \caption{(Color online) Charge density variation $\Delta n_{i,\sigma}$ as a function of distance $|R_i|$ to the impurity along the armchair direction. (a) and (b) are the cases of finite doping, with a (a) magnetic or (b) non-magnetic Anderson impurity. (c) and (d) are the half filling cases. The impurity is in magnetic phase in (c) and nonmagnetic phase in (d). The parameters  ($\varepsilon_0$,$U$,$V$,$E_f$) : (a) (-0.65, 3, 1, 0.6) eV; (b) (-1.2, 1, 0.6, 0.6) eV; (c) (-0.96, 2.4, 3, 0) eV; (d)(-1.19, 3, 5.8, 0) eV.
(e)-(h) are contour plots of $\Delta n_{i,\sigma}$ of (a) in graphene plane, where impurity is at the origin and the black dashed lines indicate armchair direction.}
 \label{fig1}
\end{figure*}

With the calculated $\langle n_\sigma \rangle$, the retarded real-space Green's function of graphene  can then be  calculated with the Dyson equation
\begin{equation}\label{realgreen}
  G_{ii,\sigma}(\omega)=\textrm{g}_{ii}(\omega)+\textrm{g}_{i0}(\omega) T_{00,\sigma}(\omega) \textrm{g}_{0i}(\omega),
\end{equation}
where the the influence of the impurity is attributed to a spin dependent diagonal element of the $T$-matrix,
$T_{00,\sigma}(\omega)=V^2/[\omega-\varepsilon_\sigma-V^2\textrm{g}_{00}(\omega)+i0^{\dag}]$.
In principle, the Anderson impurity can be equivalently viewed as a
 spin dependent local potential $V^2/(\omega-\varepsilon_\sigma+i0^{\dag})$ applied on the host carbon atom~\cite{adsorcond}.

Now, with the Green's function in Eq.~\eqref{realgreen},  we can calculate other quantities. The variation of the charge density due to  the impurity scattering is given by
\begin{equation}\label{chardensity}
  \Delta n_{i,\sigma} = -\frac{1}{\pi}\textmd{Im} \int^{\infty}_{-\infty}d\omega f(\omega)\Delta G_{ii,\sigma}(\omega),
\end{equation}
where $\Delta G_{ii,\sigma}(\omega)= G_{ii,\sigma}(\omega)-\textrm{g}_{ii}(\omega)$ is the variation of the retarded Green's function.
Since  spin  degree  of  freedom  is  considered,  the corresponding local magnetic moment on each site can be evaluated by
\begin{equation}\label{denmoment}
  M_i=[\Delta n_{i,\uparrow} -\Delta n_{i,\downarrow}] \mu_B.
\end{equation}
The change of the real-space LDOS is
\begin{equation}\label{ldos}
  \Delta \rho_{i,\sigma}(\omega)=-\frac{1}{\pi}\textmd{Im} \Delta G_{ii,\sigma}(\omega).
\end{equation}
By Fourier transformation, we get the FT-LDOS~\cite{FOproquasi}
\begin{equation}\label{ftldos}
 \begin{split}
  \Delta\rho_{\sigma}(q,\omega)=\frac{i}{2\pi N} \sum_{\alpha,k}
  [ &\Delta G_{\alpha\alpha,\sigma}(k+q,k,\omega)- \\
    &\Delta G^{\ast}_{\alpha\alpha,\sigma}(k,k+q,\omega)], 
 \end{split}
\end{equation}
where $N$ is the number of unit cells, $\alpha=\{A, B\}$ is sublattice index and $\Delta G_{\alpha\alpha,\sigma}(k',k,\omega)$ is the variation of retarded Green's function in momentum space. With all the formulae above, we can numerically calculate the spin-dependent FO and QPI pattern in graphene in the presence of an Anderson impurity.

\section{Results and discussions}\label{results}
Here, we will focus on the cases when the Anderson impurity is in magnetic phase, since that the cases of a nonmagnetic impurity has been intensively studied~\cite{effmonobi,ldosgraphite,FOImScaTemp,sublatticeasy,shortspapattern,FONlayer}.
We first discuss the spin dependent FO in graphene induced by an Anderson impurity.
We then give an analytic  formula to describe this spin dependent FO and local magnetic moment oscillation. Finally, we calculate the spin dependent FT-LDOS to understand the scattering and interference of quasiparticles of an magnetic Anderson impurity.

\subsection{Numerical results about the FOs}
Anderson impurity on graphene has been intensively studied in last decade, and its magnetic phase diagram is well known~\cite{PRL2008,zqzhangprb}.  Here, we use the mean field approximation and do not consider the Kondo physics~\cite{tunkondgrap,kondadagrap,xieKondo,kondmaggrap,orbtunkondgrap}. It should be noted that to observe the Kondo physics on graphene is still a big challenge to experiment, and so far most of the experiments about the adatoms on graphene can be understood at the mean field level. In our numerical calculations,  by choosing proper values of the parameters~(\textit{i}.\textit{e}., $\varepsilon_0$, $U$, $E_f$, and $V$), we can study the FOs for both the magnetic and nonmagnetic phases of the Anderson impurity.
For example, $\varepsilon_0=-0.65$ eV, $U=3$ eV, $V=1$ eV, $E_f=0.6$ eV are the typical values of parameters for magnetic impurity.

In addition to the spin degree of freedom, the  two sublattices should be treated separately. It is because that the top site impurity breaks the sublattice symmetry.  It is  indicated that the FOs on the two sublattices are different in the presence of potential scatters~\cite{sublatticeasy}.

We first show the spin resolved charge density variation $\Delta n_{i,\sigma}$ on sublattices A and B in Fig.~\ref{fig1}, where the charge density is plotted along a line in the armchair direction, and the horizontal axis $|R_i|$ is the distance to the Anderson impurity. Figs.~\hyperref[fig1]{1(a)} and \hyperref[fig1]{1(b)} are for the doping case with Fermi energy $E_f$=0.6 eV.  We see that, when the impurity becomes magnetic [Fig.~\hyperref[fig1]{1(a)}], the FOs for the up and down spin become different, and they have opposite sign  on the same site. So, in Fig.~\hyperref[fig1]{1(a)}, the charge density oscillations have both sublattice and spin asymmetry.  In order to show the spin asymmetry more clearly, we plot $\Delta n_{i,\sigma}$ in the graphene plane (impurity is at the origin) in Figs.~\hyperref[fig1]{1(e)}-\hyperref[fig1]{1(h)}. For example, it is clear that, on sublattice A, $\Delta n_{i,\sigma}$ for up spin [Fig.~\hyperref[fig1]{1(e)}] and down spin [Fig.~\hyperref[fig1]{1(f)}] behave oppositely. The case of sublattice B is given in Figs.~\hyperref[fig1]{1(g)} and \hyperref[fig1]{1(h)}.
Later, we will give an analytic formula about this spin dependent FO. Actually, the charge density for the up and down spin both oscillate as a sine function, but have a phase shift. With the parameters used in Fig.~\hyperref[fig1]{1(a)}, the phase shift induces a minus, so that the charge density variations for up and down spin on the same site always have opposite sign.
As a comparison, when the Anderson impurity is in nonmagnetic phase [Fig.~\hyperref[fig1]{1(b)}], the FO only has sublattice asymmetry, which is in agreement with former literature~\cite{sublatticeasy}.
In  addition, for the doping case, the amplitude of FO  always decays as $r^{-2}$, no matter whether the impurity is magnetic or not.  This also can be shown clearly by the analytic formula  given in next section.

In Figs.~\hyperref[fig1]{1(c)} and \hyperref[fig1]{1(d)},  we plot the charge  density variation at half filling. Since $k_f=0$, $\Delta n_{i,\sigma}$ will not oscillate anymore, but decay as $r^{-3}$. When the impurity is magnetic [Fig.~\hyperref[fig1]{1(c)}], the spin and sublattice asymmetry still occur. And, in the nonmagnetic case  [Fig.~\hyperref[fig1]{1(d)}], only sublattice asymmetry can be observed.

\begin{figure}
\centering
\includegraphics[width=8.6cm]{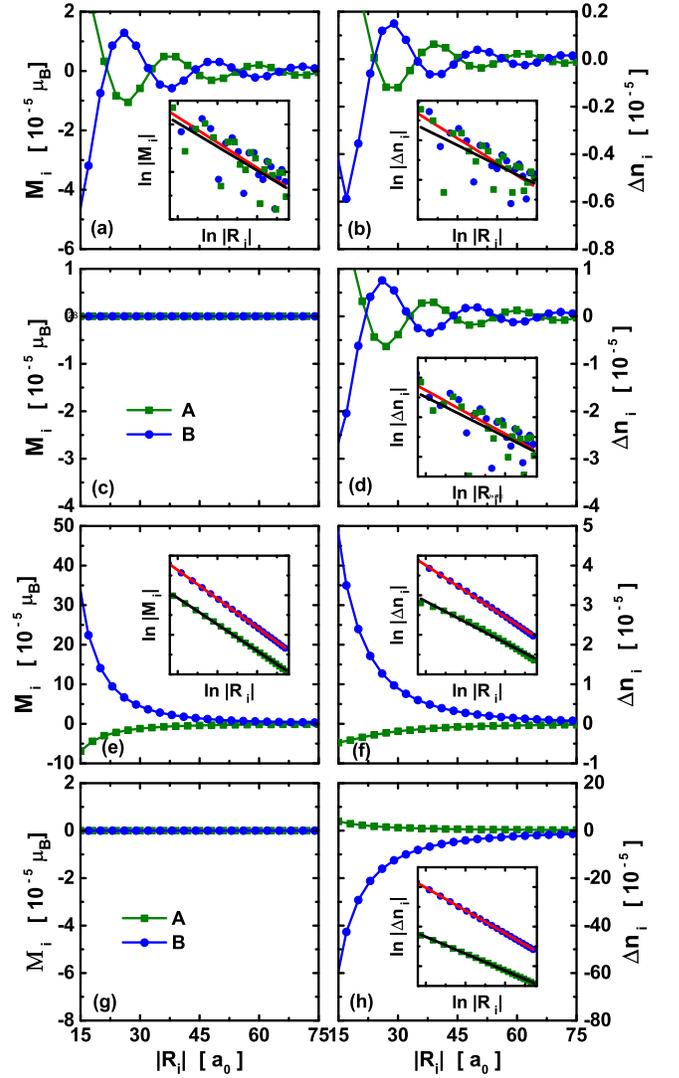}
\caption{(Color online) Local magnetic moment $M_i$ (left column) and total charge density (right column) as functions of distance $|R_i|$ to the impurity along the armchair direction.
(a) and (b) correspond to the case in Fig.~\hyperref[fig1]{1(a)}; (c) and (d) to Fig.~\hyperref[fig1]{1(b)}; (e) and (f) to Fig.~\hyperref[fig1]{1(c)}; (g) and (h) to Fig.~\hyperref[fig1]{1(d)}.
 Insets: $\log |\Delta n_i|$ (or $\log|M_i|$) on sublattices A and B as functions of $\log |R_i|$. Black and red lines are linear fittings to the log-log plots.}
\label{fig2}
\end{figure}

This spin dependent FO, together with the sublattice asymmetry, will give rise to  some interesting phenomena. One is the appearance of local magnetic moment oscillation. As shown in Fig.~\hyperref[fig1]{1(a)}, the FO of electron density with up spin has a phase shift relative to that with down spin. It implies that the local magnetic moment on graphene will become nonzero, when the Anderson impurity is magnetic.
Meanwhile, this spin asymmetry will also suppress the oscillation amplitude of total charge density, since that $\Delta n_{i,\sigma}$ for spin up and down may cancel each other out. We plot the corresponding local magnetic moment $M_i=(\Delta n_{i\uparrow}-\Delta n_{i\downarrow})\mu_B$ and total charge density variation $\Delta n_i = \Delta n_{i\uparrow}+ \Delta n_{i\downarrow}$ in Fig.~\ref{fig2}. With finite doping, an obvious local magnetic moment oscillation is found when the impurity is magnetic [see Fig.~\hyperref[fig2]{2(a)}], and the corresponding charge density oscillation is given in  Fig.~\hyperref[fig2]{2(b)}. Note that, both the charge and local magnetic moment oscillations are sublattice asymmetric. If the impurity is in nonmagnetic phase, the spin degree is degenerate and local magnetic moment becomes zero [see Fig.~\hyperref[fig2]{2(c)}]. In this case, only charge density oscillation can be found [see Fig.~\hyperref[fig2]{2(d)}].

Figs.~\hyperref[fig2]{2(e)}-\hyperref[fig2]{2(h)} are the results for the half filling case. Though there is no oscillation at half filling,  local magnetic moment decay still occur for a magnetic Anderson impurity [see Fig.~\hyperref[fig2]{2(e)}]. Note that, the local magnetic moment on sublattice A is  very tiny, and an obvious decay of local magnetic moment can only be found on sublattice B. Interestingly, this sublattice dependent magnetic moment decay is very like the one observed in the recent STM experiment~\cite{atomcontr,Li2019}, where obvious local magnetic moment decay is only found on sublattice B around a hydrogen impurity.
The corresponding charge density variation $\Delta n_i$ is also given in Fig.~\hyperref[fig2]{2(f)}. If the impurity is nonmagnetic, the local magnetic moments on lattice sites are zero [see Fig.~\hyperref[fig2]{2(g)}], while the charge density decay  still can be found as shown in Fig.~\hyperref[fig2]{2(h)}.

To study how rapid $M_i$ and $\Delta n_i$ decay, we plot $\log |M_i|$ and $\log |\Delta n_i|$ as functions of $\log |R_i|$ in the insets of Fig.~\ref{fig2}. Here, the green squares are the data for sublattice A, and the blue dots are for sublattice B. The black and red lines are the linear fitting results for sublattice A and B, respectively.
In doped graphene [see insets of Figs.~\hyperref[fig2]{2(a)}, \hyperref[fig2]{2(b)} and \hyperref[fig2]{2(d)}], the slops of the black and red lines are about $-2$, which indicates that $\Delta n_i$ and $M_i$ both decay as $r^{-2}$.
At half filling [see insets of Figs.~\hyperref[fig2]{2(e)}, \hyperref[fig2]{2(f)} and \hyperref[fig2]{2(h)}], the slopes of the fitting lines are about $-3$, which implies a decay rate of $r^{-3}$ for both $\Delta n_i$ and $M_i$. The decay rates here are in accordance with the former studies about FO in graphene induced by local potential perturbation~\cite{effmonobi,ldosgraphite}.

Comparing the results in Fig.~\ref{fig1} and Fig.~\ref{fig2}, it is obvious that the amplitude of the total charge density oscillation is  suppressed due to the spin asymmetry of FOs.  We replot $\Delta n_{i,\sigma}$ and $\Delta n_i$ together in a concrete case and give a detailed comparison in Appendix \ref{appda}.

\subsection{Analytic formulae for the spin dependent FO}
Here, we interpret the spin dependent FO in a more analytic way with the method used in Ref.~\cite{sublatticeasy}. The starting point is the analytic expression about the retarded real-space Green's function of bare graphene
\begin{equation}\label{baregreen}
  \textmd{g}_{ij}(\omega)=\frac{A(\omega)e^{iQ(\omega)D}}{\sqrt{D}},
\end{equation}
which is derived within the stationary phase approximation (SPA)~\cite{spagrap}. Here, the subscripts $i$ and  $j$ denote two carbon sites. $D=2|R_i-R_j|/{3a_0}$ represents the separation between them with bond length $a_0=0.142$ nm as unit. $A(\omega)$ and $Q(\omega)$ depend on the sublattice configuration of sites $i$ and $j$ and direction between them.
For two sites on the same sublattice in a line along the armchair direction, we have~\cite{spagrap}
\begin{subequations}\label{aingf}
 \begin{align}
   A(\omega)&=-\sqrt{\frac{2i}{\pi}}\frac{\sqrt{\omega}}{\sqrt{(\omega^2+3t^2)\sqrt{t^2-\omega^2}}},\\
   Q(\omega)&=\textmd{arccos}(\frac{-\sqrt{t^2-\omega^2}}{t}),
 \end{align}
\end{subequations}
which are only valid in the  energy region $|\omega|<|t|$. Thus, with Eqs.~\eqref{baregreen} and \eqref{aingf}, we can get an analytic formula about the FO along the armchair direction.
Note that, this analytic bare Green's function is a rather good approximation, when  $D$ is larger than several lattice constants.

Now, with $T$-matrix formulae given in Eqs.~\eqref{realgreen} and \eqref{chardensity}, we can use this bare Green's function to get an analytic expression of the spin resolved charge density along armchair direction  at zero temperature
\begin{equation}\label{analycharden}
  \Delta n_{i,\sigma} \approx \textmd{Im} \sum^{\infty}_{n=0} \frac{\gamma_{n,\sigma}(E_f)}{D^{n+2}} e^{2i Q(E_f)D},
\end{equation}
where $E_f$ is the Fermi energy and  $D=\frac{2|R_i|}{3a_0}$ represents the distance from the impurity.
The sum coefficients $\gamma_{n,\sigma}(E_f)$ is defined as
\begin{equation}\label{gamma}
  \gamma_{n,\sigma}(E_f)=\frac{(-1)^{n+1}L^{(n)}_{\sigma}(E_f)}{\pi[2iQ^{(1)}(E_f)]^{n+1}},
\end{equation}
where $L^{(n)}_{\sigma}(\omega)$ denotes the $\textit{n}^{th}$ derivative of the function $L_{\sigma}(\omega)=T_{00,\sigma}(\omega)A^2(\omega)$. Considering the validity of SPA, Eq.~\eqref{analycharden} can give a well description about the charge density variation for the carbon sites far away from the impurity. We also see that, when $D$ is not small, only the first several terms are needed in the calculation. The details to derive Eq.~\eqref{analycharden} are given in Appendix \ref{appb}.

Precisely speaking, Eq.~\eqref{analycharden} can only describe the FO on sublattice A in the armchair direction [see the dashed black lines in Figs.~\hyperref[fig1]{1(e)} and \hyperref[fig1]{1(f)}]. But due to the rotation symmetry of the graphene lattice [see Figs.~\hyperref[fig1]{1(e)} and \hyperref[fig1]{1(f)}], it can also approximately describe the FO along other directions. However, we can not give an concrete formula about the FO on sublattice B with similar method, because that we do not have simple expressions about $A(\omega)$ and $Q(\omega)$ when $i$, $j$ belong to different sublattices.

Next, we discuss the characteristics of FO described by Eq.~\eqref{analycharden}.
In undoped system with $E_f \rightarrow 0$, we have $Q(E_f) \rightarrow 0$.
And the first two terms of  $\gamma_{n,\sigma}(E_f)$ ($n=0,1$) are
\begin{subequations}
 \begin{align}
   \lim_{E_f \rightarrow 0}\gamma_{0,\sigma}(E_f) &=0; \\
   \lim_{E_f \rightarrow 0}\gamma_{1,\sigma}(E_f) &=\frac{i V^2}{6\pi^2 t} \frac{-1}{\varepsilon_\sigma-i 0^{\dag}}.
 \end{align}
\end{subequations}
Considering the first non-zero term in Eq.~\eqref{analycharden},
the charge density can be approximated as
\begin{equation}\label{neutcharden}
  \Delta n_{i,\sigma} \approx -\frac{1}{\varepsilon_\sigma} \frac{V^2}{6\pi^2 t} \frac{1}{D^3}.
\end{equation}
It means that, at half filling, there is no  oscillation, but the  charge density decreases monotonically as $r^{-3}$. More importantly, if the Anderson impurity is magnetic and
$\varepsilon_{\uparrow}<0<\varepsilon_{\downarrow}$,   $\Delta n_{i,\uparrow}$ and $\Delta n_{i,\downarrow}$ always have opposite sign. This is just the situation in Fig.~\hyperref[fig1]{1(c)}.
The corresponding local magnetic moment on sublattice A is  given by
\begin{equation}
 M_i  \approx (\frac{1}{\varepsilon_\uparrow}-\frac{1}{\varepsilon_\downarrow}) \frac{-V^2 \mu_B}{6\pi^2 t}  \frac{1}{D^3},
\end{equation}
which indicates a $r^{-3}$ decay, and is consist with the numerical results.
Meanwhile, when the Anderson impurity is nonmagnetic with $\epsilon_{\uparrow}=\epsilon_{\downarrow}$, we have $\Delta n_{i,\uparrow}=\Delta n_{i,\downarrow}$ and thus $M_i=0$.
Compared with the numerical results, the analytic formulae above work quite well for the carbon atoms far away from the impurity in the half filling case.

\begin{figure}
\centering
\includegraphics[width=8.6cm]{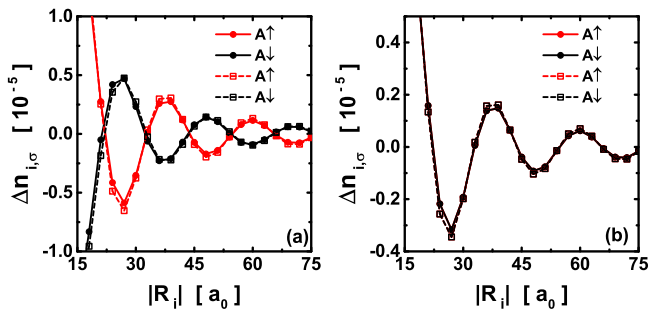}
\caption{(Color online) Comparison of $\Delta n_{i,\sigma}$ on sublattice A between numerical results (solid lines with filled circles) got by $T$-matrix method and analytic results (dashed lines with empty squares) calculated by formula \eqref{dopecharden}. (a) corresponds to the case of Fig.~\hyperref[fig1]{1(a)},  and (b) to Fig.~\hyperref[fig1]{1(b)}.}
\label{fig3}
\end{figure}

For finite doping, $Q(E_f) \neq 0$ and the first nonzero sum coefficient is
\begin{equation}
  \gamma_{0,\sigma}(E_f)=\frac{-L_{\sigma}(E_f)}{2\pi iQ^{(1)}(E_f)} \neq 0.
\end{equation}
Ignoring other terms,  we have
\begin{equation}\label{dns}
 \Delta n_{i,\sigma} \approx \textmd{Im}\ \frac{\gamma_{0,\sigma}(E_f)}{D^2}e^{2iQ(E_f)D}.
\end{equation}
Considering that $\gamma_{0,\sigma}(E_f)$ is complex valued, the above equation can be recast as
\begin{equation} \label{charden}
  \Delta n_{i,\sigma} \approx \frac{|\gamma_{0,\sigma}(E_f)|}{D^2} \sin[2Q(E_f)D + \theta^{\gamma}_{0,\sigma}(E_f)],
\end{equation}
where $\theta^{\gamma}_{0,\sigma}(E_f)$ is the phase of $\gamma_{0,\sigma}(E_f)$.
A simpler expression can be got when $|E_f| < |t|$ and  $|E_f-\varepsilon_\sigma | \gg |V^2 \textmd{g}_{00}(E_f)|$. In this situation, $\theta^{\gamma}_{0,\sigma}(E_f)$ is $0$ or $\pi$ depending on the sign of $\varepsilon_\sigma - E_f$.
Then, $\Delta n_{i,\sigma}$ reduces to
\begin{equation}\label{dopecharden}
 \Delta n_{i,\sigma} \approx \frac{1}{\varepsilon_\sigma-E_f} \frac{V^2 |E_f| }{\pi^2 (E^2_f+3t^2) } \frac{\sin[2Q(E_f)D]}{D^2}.
\end{equation}
Interestingly, Eq.~\eqref{dopecharden} is just the case shown in Figs.~\hyperref[fig1]{1(a)} and \hyperref[fig1]{1(b)}. In Fig.~\hyperref[fig1]{1(a)},  according to Eq.~\eqref{dopecharden},  the charge density variations for up spin and down spin on the same site always have opposite sign  because  $\varepsilon_\uparrow < E_f < \varepsilon_\downarrow$, so that  the FOs become spin asymmetric. And the corresponding magnetic moment is given by
\begin{small}
\begin{equation}
  M_i \approx  \left( \frac{1}{\varepsilon_\uparrow-E_f}-\frac{1}{\varepsilon_\downarrow-E_f} \right) \frac{V^2 |E_f| \mu_B}{\pi^2 (E^2_f+3t^2) } \frac{\sin[2Q(E_f)D]}{D^2}.
\end{equation}
\end{small}
Here, we see that both $\Delta n_{i,\sigma}$ and $M_i$ on sublattice A will oscillate as a sine function with envelope decaying like $r^{-2}$.
In the nonmagnetic case [Fig.~\hyperref[fig1]{1(b)}] where $\varepsilon_\uparrow = \varepsilon_\downarrow$, $\Delta n_{i,\uparrow}=\Delta n_{i,\downarrow}$ and $M_i=0$.

Another useful message is the period of the FO. Eq.~\eqref{charden} indicates that the period should be
\begin{equation}
  T_{p}=\frac{3\pi a_0}{2Q(E_f)}.
\end{equation}
For example, when $E_f=0.6$ eV, the oscillation period is about $22a_0$, which is in agreement with the numerical results in Fig.~\ref{fig1}. Actually, the oscillation period depends on the Fermi energy $E_f$, \textit{i}.\textit{e}. Fermi momentum $k_f$. The larger $k_f$ is, the shorter the oscillation period is.

The behaviors of FOs revealed by the analytic expression \eqref{dopecharden} agrees well with the numerical results in last section. In Fig.~\ref{fig3}, we give a comparison between the analytic formula \eqref{dopecharden} and the numerical results got in last section. In Fig.~\hyperref[fig3]{3(a)}, we plot $\Delta n_{i,\sigma}$ for the magnetic case with finite doping as in Fig.~\hyperref[fig1]{1(a)}.  The  solid lines with filled circles  represent the charge density variations on  sublattice A got by numerical calculations, where the red (black) ones denote up spin (down spin). The dashed lines with empty squares represent $\Delta n_{i,\sigma}$ got by Eq.~\eqref{dopecharden}, where red (black) ones are for up (down) spin.
 As shown in Fig.~\hyperref[fig3]{3(a)}, although some approximations are used, the analytic formula works quite well.  The comparison for the nonmagnetic case corresponding to Fig.~\hyperref[fig1]{1(b)} is given in Fig.~\hyperref[fig3]{3(b)}. Note that the analytical formulae above only work well for the cases when the carbon atoms  are far away from the impurity.

\subsection{Spin dependent FT-LDOS}
Besides the charge density, the energy dependent LDOS is another  quantity that can be directly probed by STM.
Due to the impurity scattering, low-energy electrons (or quasiparticles) in the two valleys of graphene will interference with each other, and give rise to special LDOS patterns. Via the Fourier transformation of LDOS, \textit{i.e.}, FT-LDOS, we can acquire information about scattering,  pseudospin texture and dispersion relation of quasiparticles.

\begin{figure}
\centering
\includegraphics[width=8.6cm]{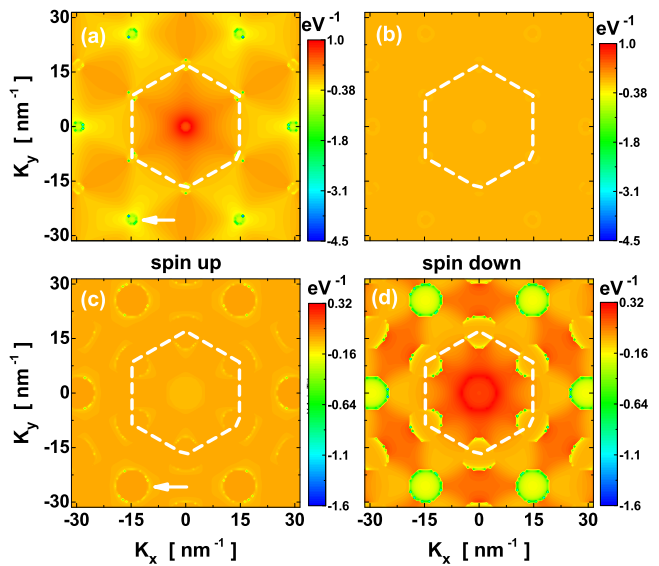}
\caption{(Color online) The real part of FT-LDOS $\Delta\rho_{\sigma}(k,\omega)$.  (a) $\omega=-0.4$ eV, spin up; (b) $\omega=-0.4$ eV, spin down; (c)  $\omega=1.3$ eV, spin up; (d)  $\omega=1.3$ eV, spin down. Other parameters: $\varepsilon_0=-0.81$ eV, $U=3$ eV, $V=2$ eV and $E_f=0.6$ eV. White dashed lines are boundary of first Brillouin zone and white arrows point to the first order spots of reciprocal lattice.}
\label{fig4}
\end{figure}


In Fig.~\ref{fig4}, we calculate the FT-LDOS $\Delta \rho_\sigma (k,\omega)$ in momentum space using Eq.~\eqref{ftldos}, where Figs.~\hyperref[fig4]{4(a)} and \hyperref[fig4]{4(b)} are at  at $\omega=-0.4$ eV and  Figs.~\hyperref[fig4]{4(c)} and \hyperref[fig4]{4(d)} are at $\omega = 1.3 $ eV. Here, the Anderson impurity is magnetic and $E_f=0.6$ eV.
The magnetic impurity makes also the FT-LDOS spin asymmetric.
For example, at $\omega=-0.4$ eV, up spin has an obvious intensity [Fig.~\hyperref[fig4]{4(a)}], while the signal of the down spin is tiny [Fig.~\hyperref[fig4]{4(b)}]. At $\omega=1.3$ eV [Figs.~\hyperref[fig4]{4(c)} and \hyperref[fig4]{4(d)}], the reverse happens.
This spin asymmetry in FT-LDOS means that the scattering of the up and down spin  electrons  occur at different energy. It is reasonable because that, when the Anderson impurity is magnetic, it can be viewed as an spin dependent local potential. Actually, in the presence of a magnetic Anderson impurity, electrons with up and down spin on graphene will form their own resonance states at different energy~\cite{Li2019,dislocgrap,resscagrap,locelegrap,adsorgrap,zqzhangprb}.

Besides the spin asymmetry, all the other typical features of the impurity induced FT-LDOS can be found in Fig.~\ref{fig4},  such as the the  high intensity region around the $\Gamma$ point, circles with $2k_f$ radius at the first order spots of the reciprocal lattice, rotational-symmetry broken rings at the corners of first Brillouin zone~\cite{quasichiralprob,pseudospingraphene,ftsts,effmonobi}. We expect that such special FT-LDOS can be detected in further spin-polarized STM experiment.

\section{Summary}\label{summary}
In summary, we theoretically investigate the Anderson impurity induced FO and QPI in graphene. We illustrate that, when the Anderson impurity is in magnetic phase, the induced FO will not only have sublattice asymmetry but also have spin asymmetry. The FO of the up and down spin electrons have a phase shift. Due to this spin asymmetry, a local magnetic moment oscillation around the impurity appears, and the total charge density oscillation will also be  modified accordingly.
Within SPA, we have also derived analytic expressions for the spin resolved charge density and the local magnetic moment to interpret the FO. We find out that, at half filling, there is no oscillation due to $k_f=0$, and the magnetic moment will decay like $1/r^3$. With finite doping, both up and down spin electrons will oscillate as a sine function but with a phase shift. Thus, the local magnetic moment oscillation is formed with the envelope decay being $1/r^2$. Finally, we numerically calculate the low-energy FT-LDOS induced by a magnetic Anderson impurity in doped graphene to analyze the QPI. It is also spin dependent, which implies that up spin and down spin electrons feel different scattering potential.


Our theory may offer some new understanding to the recent STM experiment about the hydrogen impurity on  graphene~\cite{atomcontr}. In our recent work\cite{Li2019}, we have illustrated that, for a  hydrogen impurity on graphene,  the effect of Coulomb interaction on both impurity and carbon atoms can be equivalently represented by an effective on-site $U$ of the hydrogen impurity. From this point  of view, in addition to the half filling case, it is reasonable to expect that local magnetic moment oscillation on both sublattice can be found around the hydrogen adatom with finite doping. We expect that such prediction can be tested by future  spin-polarized STM experiment.

\begin{acknowledgements}
S.L. and J.H.G. are supported by the National Key Research and Development Program of China (Grants No. 2017YFA0403501 and No. 2016YFA0401003), and National Natural Science Foundation of China (Grants No. 11534001, 11874160, 11274129).
\end{acknowledgements}

\begin{appendices}
\renewcommand{\theequation}{S\arabic{equation}}
\renewcommand{\thefigure}{S\arabic{figure}}
\setcounter{figure}{0}
\setcounter{equation}{0}

\appendix
\section{Suppression of total charge density oscillation in doped graphene by magnetic Anderson impurity}\label{appda}

Here, we illustrate that in doped graphene, once the Anderson impurity is in magnetic phase, the amplitude of total charge density oscillation can possibly be suppressed, in addition to the appearance of the local magnetic moment oscillation around the impurity. It is because that, with the parameters used here, the FO for up and down spin  have a phase shift of about $\pi$. Thus, the charge density oscillation is suppressed since the FO of the two spin cancel each other.

\begin{figure}[ht]
\centering
\includegraphics[width=8.6cm]{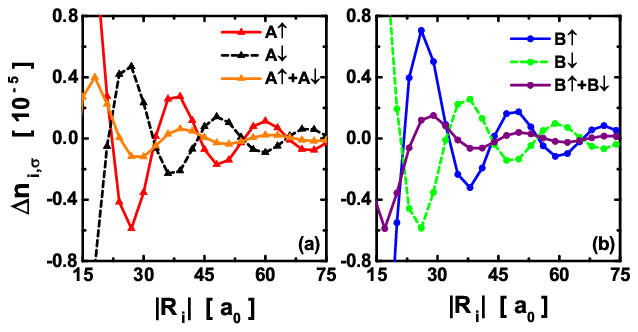}
\caption{(Color online) Replots of $\Delta n_{i,\sigma}$ and $\Delta n_i=\sum_{\sigma} \Delta n_{i,\sigma}$ for the case of Fig.~\hyperref[fig1]{1(a)}. (a) is for sublattice A and (b) for sublattice B.}
\label{figs1}
\end{figure}

To clarify this, we replot the charge density of both spin up and down [Fig.~\hyperref[fig1]{1(a)}] and total charge density [Fig.~\hyperref[fig2]{2(b)}] in Fig.~\ref{figs1}. It is obvious that the amplitude of total charge density oscillation is much smaller than that of each spin.

\section{Analytic expression for the spin dependent charge density}\label{appb}
We give the detailed  derivation about the charge density variation $\Delta n_{i,\sigma}$.
Here, the site $i$ belongs to the sublattice the impurity is coupled to, for which an analytic formula can be given.
Starting with Eq.~\eqref{chardensity} in the main
\onecolumngrid
\noindent
text, the charge density variation is given by
\begin{equation}
 \begin{split}
  \Delta n_{i,\sigma} &= -\frac{1}{\pi}\textmd{Im} \int_{-\infty}^{\infty}d\omega f(\omega)\Delta G_{ii,\sigma}(\omega) \\
  &= -\frac{1}{\pi}\textmd{Im} \int_{-\infty}^{\infty}d\omega f(\omega) \textrm{g}^2_{i0}(\omega) T_{00,\sigma}.
 \end{split}
\end{equation}
Substituting $\textmd{g}_{i0}(\omega)$ in Eq.~\eqref{baregreen} into above, we have
\begin{equation}\label{equA2}
  \Delta n_{i,\sigma} = -\frac{1}{\pi}\textmd{Im} \int_{-\infty}^{\infty}d\omega \frac{L_{\sigma}(\omega)e^{2iQ(\omega)D}}{D[1+\exp(\frac{\omega-\mu}{k_B T})]},
\end{equation}
where  $L_{\sigma}(\omega)=T_{00,\sigma}(\omega) A^2 (\omega)$, $D=\frac{2|R_i|}{3a_0}$, $\mu$ is chemical potential and $T$ denotes the temperature. The integral here can be done with the residue theorem.
The singularities of the integrand are $\omega_p=\mu+i(2p+1)\pi k_B T \quad (p=0,1,\cdots)$, at which the residue is given by
\begin{equation}
    Res(\omega_p) = \mathop{\textmd{lim}}\limits_{\omega \rightarrow \omega_p} \frac{(\omega-\omega_p)L_{\sigma}(\omega)e^{2iQ(\omega)D}}{D[1+\exp(\frac{\omega-\mu}{k_BT})]}
    = -\frac{k_BT}{D}L_{\sigma}(\omega_p)e^{2iQ(\omega_p)D}.
\end{equation}
Then, Eq.~\eqref{equA2} reduces to
\begin{equation}
   \Delta n_{i,\sigma} = -\frac{1}{\pi}\textmd{Im}\ 2\pi i \sum^{\infty}_{p=0}Res(\omega_p)
   =\textmd{Im}\ \frac{2k_BTi}{D} \sum^{\infty}_{p=0}L_{\sigma}(\omega_p)e^{2iQ(\omega_p)D}.
\end{equation}
We next expand $L_{\sigma}(\omega_p)$ and $Q(\omega_p)$ in Taylor series around $\mu$, where $Q(\omega_p)$ is only expanded to the second order.
Then, we have
\begin{small}
\begin{equation}
 \begin{split}
  \Delta n_{i,\sigma} &\approx  \textmd{Im}\ \frac{2k_BTi}{D} \sum_{n=0}^{\infty} \frac{L_{\sigma}^{(n)}(\mu)}{n!} e^{2iQ(\mu)D}
    \sum_{p=0}^{\infty} e^{2iQ^{(1)}(\mu)(\omega_p-\mu)D} (\omega_p-\mu)^n \\
    &\approx  \textmd{Im}\ \frac{2k_BTi}{D} \sum_{n=0}^{\infty} \frac{L_{\sigma}^{(n)}(\mu)}{n!} \frac{e^{2iQ(\mu)D} }{[2iQ^{(1)}(\mu)]^n}
    \frac{d^n }{d D^n} \left\{ \sum_{p=0}^{\infty} e^{2i Q^{(1)}(\mu)(\omega_p-\mu)D} \right\} \\
    &\approx  \textmd{Im}\ \frac{2i}{D} \sum_{n=0}^{\infty}  \frac{L_{\sigma}^{(n)}(\mu)}{n!} \frac{e^{2iQ(\mu)D} }{[2iQ^{(1)}(\mu)]^n}
    \frac{d^n}{d D^n} \left\{ \frac{k_BT}{2\sinh[2Q^{(1)}(\mu)\pi k_B T D]} \right\}
 \end{split}
\end{equation}
\end{small}
Taking the limit $ T\rightarrow 0$ K,
\begin{equation}
    \Delta n_{i,\sigma} \approx \textmd{Im}\ \sum_{n=0}^{\infty} \frac{(-1)^{n+1} L_{\sigma}^{(n)}(E_f)}{\pi[2iQ^{(1)}(E_f)]^{n+1}} \frac{e^{2iQ(E_f)D}}{D^{n+2}}
    \approx \textmd{Im}\ \sum_{n=0}^{\infty} \frac{\gamma_{n,\sigma}(E_f)}{D^{n+2}} e^{2iQ(E_f)D}.
\end{equation}
We get  the analytic expression for the charge density variation, which is just Eq.~\eqref{analycharden} in the main text.

\end{appendices}

\twocolumngrid
\bibliographystyle{apsrev4-1}
\bibliography{FOs}

\begin{thebibliography}{51}%
\makeatletter
\providecommand \@ifxundefined [1]{%
 \@ifx{#1\undefined}
}%
\providecommand \@ifnum [1]{%
 \ifnum #1\expandafter \@firstoftwo
 \else \expandafter \@secondoftwo
 \fi
}%
\providecommand \@ifx [1]{%
 \ifx #1\expandafter \@firstoftwo
 \else \expandafter \@secondoftwo
 \fi
}%
\providecommand \natexlab [1]{#1}%
\providecommand \enquote  [1]{``#1''}%
\providecommand \bibnamefont  [1]{#1}%
\providecommand \bibfnamefont [1]{#1}%
\providecommand \citenamefont [1]{#1}%
\providecommand \href@noop [0]{\@secondoftwo}%
\providecommand \href [0]{\begingroup \@sanitize@url \@href}%
\providecommand \@href[1]{\@@startlink{#1}\@@href}%
\providecommand \@@href[1]{\endgroup#1\@@endlink}%
\providecommand \@sanitize@url [0]{\catcode `\\12\catcode `\$12\catcode
  `\&12\catcode `\#12\catcode `\^12\catcode `\_12\catcode `\%12\relax}%
\providecommand \@@startlink[1]{}%
\providecommand \@@endlink[0]{}%
\providecommand \url  [0]{\begingroup\@sanitize@url \@url }%
\providecommand \@url [1]{\endgroup\@href {#1}{\urlprefix }}%
\providecommand \urlprefix  [0]{URL }%
\providecommand \Eprint [0]{\href }%
\providecommand \doibase [0]{http://dx.doi.org/}%
\providecommand \selectlanguage [0]{\@gobble}%
\providecommand \bibinfo  [0]{\@secondoftwo}%
\providecommand \bibfield  [0]{\@secondoftwo}%
\providecommand \translation [1]{[#1]}%
\providecommand \BibitemOpen [0]{}%
\providecommand \bibitemStop [0]{}%
\providecommand \bibitemNoStop [0]{.\EOS\space}%
\providecommand \EOS [0]{\spacefactor3000\relax}%
\providecommand \BibitemShut  [1]{\csname bibitem#1\endcsname}%
\let\auto@bib@innerbib\@empty
\bibitem [{\citenamefont {Friedel}(1952)}]{Friedel}%
  \BibitemOpen
  \bibfield  {author} {\bibinfo {author} {\bibfnamefont {J.}~\bibnamefont
  {Friedel}},\ }\href {\doibase 10.1080/14786440208561086} {\bibfield
  {journal} {\bibinfo  {journal} {Philos. Mag.}\ }\textbf {\bibinfo {volume}
  {43}},\ \bibinfo {pages} {153} (\bibinfo {year} {1952})}\BibitemShut
  {NoStop}%
\bibitem [{\citenamefont {Crommie}\ \emph {et~al.}(1993)\citenamefont
  {Crommie}, \citenamefont {Lutz},\ and\ \citenamefont
  {Eigler}}]{ImageStandWave}%
  \BibitemOpen
  \bibfield  {author} {\bibinfo {author} {\bibfnamefont {M.~F.}\ \bibnamefont
  {Crommie}}, \bibinfo {author} {\bibfnamefont {C.~P.}\ \bibnamefont {Lutz}}, \
  and\ \bibinfo {author} {\bibfnamefont {D.~M.}\ \bibnamefont {Eigler}},\
  }\href {\doibase 10.1038/363524a0} {\bibfield  {journal} {\bibinfo  {journal}
  {Nature}\ }\textbf {\bibinfo {volume} {363}},\ \bibinfo {pages} {524}
  (\bibinfo {year} {1993})}\BibitemShut {NoStop}%
\bibitem [{\citenamefont {Hasegawa}\ and\ \citenamefont
  {Avouris}(1993)}]{STMSurStep}%
  \BibitemOpen
  \bibfield  {author} {\bibinfo {author} {\bibfnamefont {Y.}~\bibnamefont
  {Hasegawa}}\ and\ \bibinfo {author} {\bibfnamefont {P.}~\bibnamefont
  {Avouris}},\ }\href {\doibase 10.1103/PhysRevLett.71.1071} {\bibfield
  {journal} {\bibinfo  {journal} {Phys. Rev. Lett.}\ }\textbf {\bibinfo
  {volume} {71}},\ \bibinfo {pages} {1071} (\bibinfo {year}
  {1993})}\BibitemShut {NoStop}%
\bibitem [{\citenamefont {Petersen}\ \emph {et~al.}(1998)\citenamefont
  {Petersen}, \citenamefont {Sprunger}, \citenamefont {Hofmann}, \citenamefont
  {L\ae{}gsgaard}, \citenamefont {Briner}, \citenamefont {Doering},
  \citenamefont {Rust}, \citenamefont {Bradshaw}, \citenamefont {Besenbacher},\
  and\ \citenamefont {Plummer}}]{FerConFTSTM}%
  \BibitemOpen
  \bibfield  {author} {\bibinfo {author} {\bibfnamefont {L.}~\bibnamefont
  {Petersen}}, \bibinfo {author} {\bibfnamefont {P.~T.}\ \bibnamefont
  {Sprunger}}, \bibinfo {author} {\bibfnamefont {P.}~\bibnamefont {Hofmann}},
  \bibinfo {author} {\bibfnamefont {E.}~\bibnamefont {L\ae{}gsgaard}}, \bibinfo
  {author} {\bibfnamefont {B.~G.}\ \bibnamefont {Briner}}, \bibinfo {author}
  {\bibfnamefont {M.}~\bibnamefont {Doering}}, \bibinfo {author} {\bibfnamefont
  {H.-P.}\ \bibnamefont {Rust}}, \bibinfo {author} {\bibfnamefont {A.~M.}\
  \bibnamefont {Bradshaw}}, \bibinfo {author} {\bibfnamefont {F.}~\bibnamefont
  {Besenbacher}}, \ and\ \bibinfo {author} {\bibfnamefont {E.~W.}\ \bibnamefont
  {Plummer}},\ }\href {\doibase 10.1103/PhysRevB.57.R6858} {\bibfield
  {journal} {\bibinfo  {journal} {Phys. Rev. B}\ }\textbf {\bibinfo {volume}
  {57}},\ \bibinfo {pages} {R6858} (\bibinfo {year} {1998})}\BibitemShut
  {NoStop}%
\bibitem [{\citenamefont {Dalla~Torre}\ \emph
  {et~al.}(2016{\natexlab{a}})\citenamefont {Dalla~Torre}, \citenamefont {He},\
  and\ \citenamefont {Demler}}]{Holomap}%
  \BibitemOpen
  \bibfield  {author} {\bibinfo {author} {\bibfnamefont {E.~G.}\ \bibnamefont
  {Dalla~Torre}}, \bibinfo {author} {\bibfnamefont {Y.}~\bibnamefont {He}}, \
  and\ \bibinfo {author} {\bibfnamefont {E.}~\bibnamefont {Demler}},\ }\href
  {\doibase 10.1038/nphys3829} {\bibfield  {journal} {\bibinfo  {journal} {Nat.
  Phys.}\ }\textbf {\bibinfo {volume} {12}},\ \bibinfo {pages} {1052} (\bibinfo
  {year} {2016}{\natexlab{a}})}\BibitemShut {NoStop}%
\bibitem [{\citenamefont {Vishik}\ \emph {et~al.}(2009)\citenamefont {Vishik},
  \citenamefont {Nowadnick}, \citenamefont {Lee}, \citenamefont {Shen},
  \citenamefont {Moritz}, \citenamefont {Devereaux}, \citenamefont {Tanaka},
  \citenamefont {Sasagawa},\ and\ \citenamefont {Fujii}}]{QPIStronCorr}%
  \BibitemOpen
  \bibfield  {author} {\bibinfo {author} {\bibfnamefont {I.~M.}\ \bibnamefont
  {Vishik}}, \bibinfo {author} {\bibfnamefont {E.~A.}\ \bibnamefont
  {Nowadnick}}, \bibinfo {author} {\bibfnamefont {W.~S.}\ \bibnamefont {Lee}},
  \bibinfo {author} {\bibfnamefont {Z.~X.}\ \bibnamefont {Shen}}, \bibinfo
  {author} {\bibfnamefont {B.}~\bibnamefont {Moritz}}, \bibinfo {author}
  {\bibfnamefont {T.~P.}\ \bibnamefont {Devereaux}}, \bibinfo {author}
  {\bibfnamefont {K.}~\bibnamefont {Tanaka}}, \bibinfo {author} {\bibfnamefont
  {T.}~\bibnamefont {Sasagawa}}, \ and\ \bibinfo {author} {\bibfnamefont
  {T.}~\bibnamefont {Fujii}},\ }\href {\doibase 10.1038/nphys1375} {\bibfield
  {journal} {\bibinfo  {journal} {Nat. Phys.}\ }\textbf {\bibinfo {volume}
  {5}},\ \bibinfo {pages} {718} (\bibinfo {year} {2009})}\BibitemShut {NoStop}%
\bibitem [{\citenamefont {Roushan}\ \emph {et~al.}(2009)\citenamefont
  {Roushan}, \citenamefont {Seo}, \citenamefont {Parker}, \citenamefont {Hor},
  \citenamefont {Hsieh}, \citenamefont {Qian}, \citenamefont {Richardella},
  \citenamefont {Hasan}, \citenamefont {Cava},\ and\ \citenamefont
  {Yazdani}}]{Toposurf}%
  \BibitemOpen
  \bibfield  {author} {\bibinfo {author} {\bibfnamefont {P.}~\bibnamefont
  {Roushan}}, \bibinfo {author} {\bibfnamefont {J.}~\bibnamefont {Seo}},
  \bibinfo {author} {\bibfnamefont {C.~V.}\ \bibnamefont {Parker}}, \bibinfo
  {author} {\bibfnamefont {Y.~S.}\ \bibnamefont {Hor}}, \bibinfo {author}
  {\bibfnamefont {D.}~\bibnamefont {Hsieh}}, \bibinfo {author} {\bibfnamefont
  {D.}~\bibnamefont {Qian}}, \bibinfo {author} {\bibfnamefont {A.}~\bibnamefont
  {Richardella}}, \bibinfo {author} {\bibfnamefont {M.~Z.}\ \bibnamefont
  {Hasan}}, \bibinfo {author} {\bibfnamefont {R.~J.}\ \bibnamefont {Cava}}, \
  and\ \bibinfo {author} {\bibfnamefont {A.}~\bibnamefont {Yazdani}},\ }\href
  {\doibase 10.1038/nature08308} {\bibfield  {journal} {\bibinfo  {journal}
  {Nature}\ }\textbf {\bibinfo {volume} {460}},\ \bibinfo {pages} {1106}
  (\bibinfo {year} {2009})}\BibitemShut {NoStop}%
\bibitem [{\citenamefont {Zhang}\ \emph {et~al.}(2009)\citenamefont {Zhang},
  \citenamefont {Cheng}, \citenamefont {Chen}, \citenamefont {Jia},
  \citenamefont {Ma}, \citenamefont {He}, \citenamefont {Wang}, \citenamefont
  {Zhang}, \citenamefont {Dai}, \citenamefont {Fang}, \citenamefont {Xie},\
  and\ \citenamefont {Xue}}]{TRSSurf}%
  \BibitemOpen
  \bibfield  {author} {\bibinfo {author} {\bibfnamefont {T.}~\bibnamefont
  {Zhang}}, \bibinfo {author} {\bibfnamefont {P.}~\bibnamefont {Cheng}},
  \bibinfo {author} {\bibfnamefont {X.}~\bibnamefont {Chen}}, \bibinfo {author}
  {\bibfnamefont {J.-F.}\ \bibnamefont {Jia}}, \bibinfo {author} {\bibfnamefont
  {X.}~\bibnamefont {Ma}}, \bibinfo {author} {\bibfnamefont {K.}~\bibnamefont
  {He}}, \bibinfo {author} {\bibfnamefont {L.}~\bibnamefont {Wang}}, \bibinfo
  {author} {\bibfnamefont {H.}~\bibnamefont {Zhang}}, \bibinfo {author}
  {\bibfnamefont {X.}~\bibnamefont {Dai}}, \bibinfo {author} {\bibfnamefont
  {Z.}~\bibnamefont {Fang}}, \bibinfo {author} {\bibfnamefont {X.}~\bibnamefont
  {Xie}}, \ and\ \bibinfo {author} {\bibfnamefont {Q.-K.}\ \bibnamefont
  {Xue}},\ }\href {\doibase 10.1103/PhysRevLett.103.266803} {\bibfield
  {journal} {\bibinfo  {journal} {Phys. Rev. Lett.}\ }\textbf {\bibinfo
  {volume} {103}},\ \bibinfo {pages} {266803} (\bibinfo {year}
  {2009})}\BibitemShut {NoStop}%
\bibitem [{\citenamefont {Kim}\ \emph {et~al.}(2014)\citenamefont {Kim},
  \citenamefont {Yoshizawa}, \citenamefont {Ishida}, \citenamefont {Eto},
  \citenamefont {Segawa}, \citenamefont {Ando}, \citenamefont {Shin},\ and\
  \citenamefont {Komori}}]{ProBackTI}%
  \BibitemOpen
  \bibfield  {author} {\bibinfo {author} {\bibfnamefont {S.}~\bibnamefont
  {Kim}}, \bibinfo {author} {\bibfnamefont {S.}~\bibnamefont {Yoshizawa}},
  \bibinfo {author} {\bibfnamefont {Y.}~\bibnamefont {Ishida}}, \bibinfo
  {author} {\bibfnamefont {K.}~\bibnamefont {Eto}}, \bibinfo {author}
  {\bibfnamefont {K.}~\bibnamefont {Segawa}}, \bibinfo {author} {\bibfnamefont
  {Y.}~\bibnamefont {Ando}}, \bibinfo {author} {\bibfnamefont {S.}~\bibnamefont
  {Shin}}, \ and\ \bibinfo {author} {\bibfnamefont {F.}~\bibnamefont
  {Komori}},\ }\href {\doibase 10.1103/PhysRevLett.112.136802} {\bibfield
  {journal} {\bibinfo  {journal} {Phys. Rev. Lett.}\ }\textbf {\bibinfo
  {volume} {112}},\ \bibinfo {pages} {136802} (\bibinfo {year}
  {2014})}\BibitemShut {NoStop}%
\bibitem [{\citenamefont {Mart\'{i}nez-Velarte}\ \emph
  {et~al.}(2017)\citenamefont {Mart\'{i}nez-Velarte}, \citenamefont {Kretz},
  \citenamefont {Moro-Lagares}, \citenamefont {Aguirre}, \citenamefont
  {Riedemann}, \citenamefont {Lograsso}, \citenamefont {Morell\'{o}n},
  \citenamefont {Ibarra}, \citenamefont {Garcia-Lekue},\ and\ \citenamefont
  {Serrate}}]{ChemDisordTI}%
  \BibitemOpen
  \bibfield  {author} {\bibinfo {author} {\bibfnamefont {M.~C.}\ \bibnamefont
  {Mart\'{i}nez-Velarte}}, \bibinfo {author} {\bibfnamefont {B.}~\bibnamefont
  {Kretz}}, \bibinfo {author} {\bibfnamefont {M.}~\bibnamefont {Moro-Lagares}},
  \bibinfo {author} {\bibfnamefont {M.~H.}\ \bibnamefont {Aguirre}}, \bibinfo
  {author} {\bibfnamefont {T.~M.}\ \bibnamefont {Riedemann}}, \bibinfo {author}
  {\bibfnamefont {T.~A.}\ \bibnamefont {Lograsso}}, \bibinfo {author}
  {\bibfnamefont {L.}~\bibnamefont {Morell\'{o}n}}, \bibinfo {author}
  {\bibfnamefont {M.~R.}\ \bibnamefont {Ibarra}}, \bibinfo {author}
  {\bibfnamefont {A.}~\bibnamefont {Garcia-Lekue}}, \ and\ \bibinfo {author}
  {\bibfnamefont {D.}~\bibnamefont {Serrate}},\ }\href {\doibase
  10.1021/acs.nanolett.7b00311} {\bibfield  {journal} {\bibinfo  {journal}
  {Nano Lett.}\ }\textbf {\bibinfo {volume} {17}},\ \bibinfo {pages} {4047}
  (\bibinfo {year} {2017})}\BibitemShut {NoStop}%
\bibitem [{\citenamefont {Vonau}\ \emph {et~al.}(2005)\citenamefont {Vonau},
  \citenamefont {Aubel}, \citenamefont {Gewinner}, \citenamefont {Zabrocki},
  \citenamefont {Peruchetti}, \citenamefont {Bolmont},\ and\ \citenamefont
  {Simon}}]{ErsiFTSTS}%
  \BibitemOpen
  \bibfield  {author} {\bibinfo {author} {\bibfnamefont {F.}~\bibnamefont
  {Vonau}}, \bibinfo {author} {\bibfnamefont {D.}~\bibnamefont {Aubel}},
  \bibinfo {author} {\bibfnamefont {G.}~\bibnamefont {Gewinner}}, \bibinfo
  {author} {\bibfnamefont {S.}~\bibnamefont {Zabrocki}}, \bibinfo {author}
  {\bibfnamefont {J.~C.}\ \bibnamefont {Peruchetti}}, \bibinfo {author}
  {\bibfnamefont {D.}~\bibnamefont {Bolmont}}, \ and\ \bibinfo {author}
  {\bibfnamefont {L.}~\bibnamefont {Simon}},\ }\href {\doibase
  10.1103/PhysRevLett.95.176803} {\bibfield  {journal} {\bibinfo  {journal}
  {Phys. Rev. Lett.}\ }\textbf {\bibinfo {volume} {95}},\ \bibinfo {pages}
  {176803} (\bibinfo {year} {2005})}\BibitemShut {NoStop}%
\bibitem [{\citenamefont {Chen}\ \emph {et~al.}(2017)\citenamefont {Chen},
  \citenamefont {Cheng},\ and\ \citenamefont {Wu}}]{ChenQPI}%
  \BibitemOpen
  \bibfield  {author} {\bibinfo {author} {\bibfnamefont {L.}~\bibnamefont
  {Chen}}, \bibinfo {author} {\bibfnamefont {P.}~\bibnamefont {Cheng}}, \ and\
  \bibinfo {author} {\bibfnamefont {K.}~\bibnamefont {Wu}},\ }\href
  {http://stacks.iop.org/0953-8984/29/i=10/a=103001} {\bibfield  {journal}
  {\bibinfo  {journal} {J. Phys.: Condens. Matter}\ }\textbf {\bibinfo {volume}
  {29}},\ \bibinfo {pages} {103001} (\bibinfo {year} {2017})}\BibitemShut
  {NoStop}%
\bibitem [{\citenamefont {Rutter}\ \emph {et~al.}(2007)\citenamefont {Rutter},
  \citenamefont {Crain}, \citenamefont {Guisinger}, \citenamefont {Li},
  \citenamefont {First},\ and\ \citenamefont {Stroscio}}]{scattinter}%
  \BibitemOpen
  \bibfield  {author} {\bibinfo {author} {\bibfnamefont {G.~M.}\ \bibnamefont
  {Rutter}}, \bibinfo {author} {\bibfnamefont {J.~N.}\ \bibnamefont {Crain}},
  \bibinfo {author} {\bibfnamefont {N.~P.}\ \bibnamefont {Guisinger}}, \bibinfo
  {author} {\bibfnamefont {T.}~\bibnamefont {Li}}, \bibinfo {author}
  {\bibfnamefont {P.~N.}\ \bibnamefont {First}}, \ and\ \bibinfo {author}
  {\bibfnamefont {J.~A.}\ \bibnamefont {Stroscio}},\ }\href {\doibase
  10.1126/science.1142882} {\bibfield  {journal} {\bibinfo  {journal}
  {Science}\ }\textbf {\bibinfo {volume} {317}},\ \bibinfo {pages} {219}
  (\bibinfo {year} {2007})}\BibitemShut {NoStop}%
\bibitem [{\citenamefont {Brihuega}\ \emph {et~al.}(2008)\citenamefont
  {Brihuega}, \citenamefont {Mallet}, \citenamefont {Bena}, \citenamefont
  {Bose}, \citenamefont {Michaelis}, \citenamefont {Vitali}, \citenamefont
  {Varchon}, \citenamefont {Magaud}, \citenamefont {Kern},\ and\ \citenamefont
  {Veuillen}}]{quasichiralprob}%
  \BibitemOpen
  \bibfield  {author} {\bibinfo {author} {\bibfnamefont {I.}~\bibnamefont
  {Brihuega}}, \bibinfo {author} {\bibfnamefont {P.}~\bibnamefont {Mallet}},
  \bibinfo {author} {\bibfnamefont {C.}~\bibnamefont {Bena}}, \bibinfo {author}
  {\bibfnamefont {S.}~\bibnamefont {Bose}}, \bibinfo {author} {\bibfnamefont
  {C.}~\bibnamefont {Michaelis}}, \bibinfo {author} {\bibfnamefont
  {L.}~\bibnamefont {Vitali}}, \bibinfo {author} {\bibfnamefont
  {F.}~\bibnamefont {Varchon}}, \bibinfo {author} {\bibfnamefont
  {L.}~\bibnamefont {Magaud}}, \bibinfo {author} {\bibfnamefont
  {K.}~\bibnamefont {Kern}}, \ and\ \bibinfo {author} {\bibfnamefont {J.~Y.}\
  \bibnamefont {Veuillen}},\ }\href {\doibase 10.1103/PhysRevLett.101.206802}
  {\bibfield  {journal} {\bibinfo  {journal} {Phys. Rev. Lett.}\ }\textbf
  {\bibinfo {volume} {101}},\ \bibinfo {pages} {206802} (\bibinfo {year}
  {2008})}\BibitemShut {NoStop}%
\bibitem [{\citenamefont {Mallet}\ \emph {et~al.}(2012)\citenamefont {Mallet},
  \citenamefont {Brihuega}, \citenamefont {Bose}, \citenamefont {Ugeda},
  \citenamefont {G\'omez-Rodr\'{\i}guez}, \citenamefont {Kern},\ and\
  \citenamefont {Veuillen}}]{pseudospingraphene}%
  \BibitemOpen
  \bibfield  {author} {\bibinfo {author} {\bibfnamefont {P.}~\bibnamefont
  {Mallet}}, \bibinfo {author} {\bibfnamefont {I.}~\bibnamefont {Brihuega}},
  \bibinfo {author} {\bibfnamefont {S.}~\bibnamefont {Bose}}, \bibinfo {author}
  {\bibfnamefont {M.~M.}\ \bibnamefont {Ugeda}}, \bibinfo {author}
  {\bibfnamefont {J.~M.}\ \bibnamefont {G\'omez-Rodr\'{\i}guez}}, \bibinfo
  {author} {\bibfnamefont {K.}~\bibnamefont {Kern}}, \ and\ \bibinfo {author}
  {\bibfnamefont {J.~Y.}\ \bibnamefont {Veuillen}},\ }\href {\doibase
  10.1103/PhysRevB.86.045444} {\bibfield  {journal} {\bibinfo  {journal} {Phys.
  Rev. B}\ }\textbf {\bibinfo {volume} {86}},\ \bibinfo {pages} {045444}
  (\bibinfo {year} {2012})}\BibitemShut {NoStop}%
\bibitem [{\citenamefont {Mallet}\ \emph {et~al.}(2016)\citenamefont {Mallet},
  \citenamefont {Brihuega}, \citenamefont {Cherkez}, \citenamefont
  {G\'{o}mez-Rodr\'{\i}guez},\ and\ \citenamefont {Veuillen}}]{foprobbystm}%
  \BibitemOpen
  \bibfield  {author} {\bibinfo {author} {\bibfnamefont {P.}~\bibnamefont
  {Mallet}}, \bibinfo {author} {\bibfnamefont {I.}~\bibnamefont {Brihuega}},
  \bibinfo {author} {\bibfnamefont {V.}~\bibnamefont {Cherkez}}, \bibinfo
  {author} {\bibfnamefont {J.~M.}\ \bibnamefont {G\'{o}mez-Rodr\'{\i}guez}}, \
  and\ \bibinfo {author} {\bibfnamefont {J.-Y.}\ \bibnamefont {Veuillen}},\
  }\href {\doibase https://doi.org/10.1016/j.crhy.2015.12.013} {\bibfield
  {journal} {\bibinfo  {journal} {C. R. Physique}\ }\textbf {\bibinfo {volume}
  {17}},\ \bibinfo {pages} {294 } (\bibinfo {year} {2016})}\BibitemShut
  {NoStop}%
\bibitem [{\citenamefont {Simon}\ \emph {et~al.}(2011)\citenamefont {Simon},
  \citenamefont {Bena}, \citenamefont {Vonau}, \citenamefont {Cranney},\ and\
  \citenamefont {Aubel}}]{ftsts}%
  \BibitemOpen
  \bibfield  {author} {\bibinfo {author} {\bibfnamefont {L.}~\bibnamefont
  {Simon}}, \bibinfo {author} {\bibfnamefont {C.}~\bibnamefont {Bena}},
  \bibinfo {author} {\bibfnamefont {F.}~\bibnamefont {Vonau}}, \bibinfo
  {author} {\bibfnamefont {M.}~\bibnamefont {Cranney}}, \ and\ \bibinfo
  {author} {\bibfnamefont {D.}~\bibnamefont {Aubel}},\ }\href
  {http://stacks.iop.org/0022-3727/44/i=46/a=464010} {\bibfield  {journal}
  {\bibinfo  {journal} {J. Phys. D: Appl. Phys.}\ }\textbf {\bibinfo {volume}
  {44}},\ \bibinfo {pages} {464010} (\bibinfo {year} {2011})}\BibitemShut
  {NoStop}%
\bibitem [{\citenamefont {Castro~Neto}\ \emph {et~al.}(2009)\citenamefont
  {Castro~Neto}, \citenamefont {Guinea}, \citenamefont {Peres}, \citenamefont
  {Novoselov},\ and\ \citenamefont {Geim}}]{eleprograp}%
  \BibitemOpen
  \bibfield  {author} {\bibinfo {author} {\bibfnamefont {A.~H.}\ \bibnamefont
  {Castro~Neto}}, \bibinfo {author} {\bibfnamefont {F.}~\bibnamefont {Guinea}},
  \bibinfo {author} {\bibfnamefont {N.~M.~R.}\ \bibnamefont {Peres}}, \bibinfo
  {author} {\bibfnamefont {K.~S.}\ \bibnamefont {Novoselov}}, \ and\ \bibinfo
  {author} {\bibfnamefont {A.~K.}\ \bibnamefont {Geim}},\ }\href {\doibase
  10.1103/RevModPhys.81.109} {\bibfield  {journal} {\bibinfo  {journal} {Rev.
  Mod. Phys.}\ }\textbf {\bibinfo {volume} {81}},\ \bibinfo {pages} {109}
  (\bibinfo {year} {2009})}\BibitemShut {NoStop}%
\bibitem [{\citenamefont {Bena}(2008)}]{effmonobi}%
  \BibitemOpen
  \bibfield  {author} {\bibinfo {author} {\bibfnamefont {C.}~\bibnamefont
  {Bena}},\ }\href {\doibase 10.1103/PhysRevLett.100.076601} {\bibfield
  {journal} {\bibinfo  {journal} {Phys. Rev. Lett.}\ }\textbf {\bibinfo
  {volume} {100}},\ \bibinfo {pages} {076601} (\bibinfo {year}
  {2008})}\BibitemShut {NoStop}%
\bibitem [{\citenamefont {Bena}\ and\ \citenamefont
  {Kivelson}(2005)}]{ldosgraphite}%
  \BibitemOpen
  \bibfield  {author} {\bibinfo {author} {\bibfnamefont {C.}~\bibnamefont
  {Bena}}\ and\ \bibinfo {author} {\bibfnamefont {S.~A.}\ \bibnamefont
  {Kivelson}},\ }\href {\doibase 10.1103/PhysRevB.72.125432} {\bibfield
  {journal} {\bibinfo  {journal} {Phys. Rev. B}\ }\textbf {\bibinfo {volume}
  {72}},\ \bibinfo {pages} {125432} (\bibinfo {year} {2005})}\BibitemShut
  {NoStop}%
\bibitem [{\citenamefont {Cheianov}\ and\ \citenamefont
  {Fal'ko}(2006)}]{FOImScaTemp}%
  \BibitemOpen
  \bibfield  {author} {\bibinfo {author} {\bibfnamefont {V.~V.}\ \bibnamefont
  {Cheianov}}\ and\ \bibinfo {author} {\bibfnamefont {V.~I.}\ \bibnamefont
  {Fal'ko}},\ }\href {\doibase 10.1103/PhysRevLett.97.226801} {\bibfield
  {journal} {\bibinfo  {journal} {Phys. Rev. Lett.}\ }\textbf {\bibinfo
  {volume} {97}},\ \bibinfo {pages} {226801} (\bibinfo {year}
  {2006})}\BibitemShut {NoStop}%
\bibitem [{\citenamefont {Lawlor}\ \emph {et~al.}(2013)\citenamefont {Lawlor},
  \citenamefont {Power},\ and\ \citenamefont {Ferreira}}]{sublatticeasy}%
  \BibitemOpen
  \bibfield  {author} {\bibinfo {author} {\bibfnamefont {J.~A.}\ \bibnamefont
  {Lawlor}}, \bibinfo {author} {\bibfnamefont {S.~R.}\ \bibnamefont {Power}}, \
  and\ \bibinfo {author} {\bibfnamefont {M.~S.}\ \bibnamefont {Ferreira}},\
  }\href {\doibase 10.1103/PhysRevB.88.205416} {\bibfield  {journal} {\bibinfo
  {journal} {Phys. Rev. B}\ }\textbf {\bibinfo {volume} {88}},\ \bibinfo
  {pages} {205416} (\bibinfo {year} {2013})}\BibitemShut {NoStop}%
\bibitem [{\citenamefont {B\'acsi}\ and\ \citenamefont
  {Virosztek}(2010)}]{shortspapattern}%
  \BibitemOpen
  \bibfield  {author} {\bibinfo {author} {\bibfnamefont {A.}~\bibnamefont
  {B\'acsi}}\ and\ \bibinfo {author} {\bibfnamefont {A.}~\bibnamefont
  {Virosztek}},\ }\href {\doibase 10.1103/PhysRevB.82.193405} {\bibfield
  {journal} {\bibinfo  {journal} {Phys. Rev. B}\ }\textbf {\bibinfo {volume}
  {82}},\ \bibinfo {pages} {193405} (\bibinfo {year} {2010})}\BibitemShut
  {NoStop}%
\bibitem [{\citenamefont {Dutreix}\ and\ \citenamefont
  {Katsnelson}(2016)}]{FONlayer}%
  \BibitemOpen
  \bibfield  {author} {\bibinfo {author} {\bibfnamefont {C.}~\bibnamefont
  {Dutreix}}\ and\ \bibinfo {author} {\bibfnamefont {M.~I.}\ \bibnamefont
  {Katsnelson}},\ }\href {\doibase 10.1103/PhysRevB.93.035413} {\bibfield
  {journal} {\bibinfo  {journal} {Phys. Rev. B}\ }\textbf {\bibinfo {volume}
  {93}},\ \bibinfo {pages} {035413} (\bibinfo {year} {2016})}\BibitemShut
  {NoStop}%
\bibitem [{\citenamefont {Anderson}(1961)}]{Andimpmod}%
  \BibitemOpen
  \bibfield  {author} {\bibinfo {author} {\bibfnamefont {P.~W.}\ \bibnamefont
  {Anderson}},\ }\href {\doibase 10.1103/PhysRev.124.41} {\bibfield  {journal}
  {\bibinfo  {journal} {Phys. Rev.}\ }\textbf {\bibinfo {volume} {124}},\
  \bibinfo {pages} {41} (\bibinfo {year} {1961})}\BibitemShut {NoStop}%
\bibitem [{\citenamefont {Uchoa}\ \emph {et~al.}(2008)\citenamefont {Uchoa},
  \citenamefont {Kotov}, \citenamefont {Peres},\ and\ \citenamefont
  {Castro~Neto}}]{PRL2008}%
  \BibitemOpen
  \bibfield  {author} {\bibinfo {author} {\bibfnamefont {B.}~\bibnamefont
  {Uchoa}}, \bibinfo {author} {\bibfnamefont {V.~N.}\ \bibnamefont {Kotov}},
  \bibinfo {author} {\bibfnamefont {N.~M.~R.}\ \bibnamefont {Peres}}, \ and\
  \bibinfo {author} {\bibfnamefont {A.~H.}\ \bibnamefont {Castro~Neto}},\
  }\href {\doibase 10.1103/PhysRevLett.101.026805} {\bibfield  {journal}
  {\bibinfo  {journal} {Phys. Rev. Lett.}\ }\textbf {\bibinfo {volume} {101}},\
  \bibinfo {pages} {026805} (\bibinfo {year} {2008})}\BibitemShut {NoStop}%
\bibitem [{\citenamefont {Zhang}\ \emph
  {et~al.}(2017{\natexlab{a}})\citenamefont {Zhang}, \citenamefont {Li},
  \citenamefont {L\"u},\ and\ \citenamefont {Gao}}]{zqzhangprb}%
  \BibitemOpen
  \bibfield  {author} {\bibinfo {author} {\bibfnamefont {Z.-Q.}\ \bibnamefont
  {Zhang}}, \bibinfo {author} {\bibfnamefont {S.}~\bibnamefont {Li}}, \bibinfo
  {author} {\bibfnamefont {J.-T.}\ \bibnamefont {L\"u}}, \ and\ \bibinfo
  {author} {\bibfnamefont {J.-H.}\ \bibnamefont {Gao}},\ }\href {\doibase
  10.1103/PhysRevB.96.075410} {\bibfield  {journal} {\bibinfo  {journal} {Phys.
  Rev. B}\ }\textbf {\bibinfo {volume} {96}},\ \bibinfo {pages} {075410}
  (\bibinfo {year} {2017}{\natexlab{a}})}\BibitemShut {NoStop}%
\bibitem [{\citenamefont {Boukhvalov}\ \emph {et~al.}(2008)\citenamefont
  {Boukhvalov}, \citenamefont {Katsnelson},\ and\ \citenamefont
  {Lichtenstein}}]{hydgraelemag}%
  \BibitemOpen
  \bibfield  {author} {\bibinfo {author} {\bibfnamefont {D.~W.}\ \bibnamefont
  {Boukhvalov}}, \bibinfo {author} {\bibfnamefont {M.~I.}\ \bibnamefont
  {Katsnelson}}, \ and\ \bibinfo {author} {\bibfnamefont {A.~I.}\ \bibnamefont
  {Lichtenstein}},\ }\href {\doibase 10.1103/PhysRevB.77.035427} {\bibfield
  {journal} {\bibinfo  {journal} {Phys. Rev. B}\ }\textbf {\bibinfo {volume}
  {77}},\ \bibinfo {pages} {035427} (\bibinfo {year} {2008})}\BibitemShut
  {NoStop}%
\bibitem [{\citenamefont {Casolo}\ \emph {et~al.}(2009)\citenamefont {Casolo},
  \citenamefont {L{\o}vvik}, \citenamefont {Martinazzo},\ and\ \citenamefont
  {Tantardini}}]{undadshydgra}%
  \BibitemOpen
  \bibfield  {author} {\bibinfo {author} {\bibfnamefont {S.}~\bibnamefont
  {Casolo}}, \bibinfo {author} {\bibfnamefont {O.~M.}\ \bibnamefont
  {L{\o}vvik}}, \bibinfo {author} {\bibfnamefont {R.}~\bibnamefont
  {Martinazzo}}, \ and\ \bibinfo {author} {\bibfnamefont {G.~F.}\ \bibnamefont
  {Tantardini}},\ }\href {\doibase 10.1063/1.3072333} {\bibfield  {journal}
  {\bibinfo  {journal} {J. Chem. Phys.}\ }\textbf {\bibinfo {volume} {130}},\
  \bibinfo {pages} {054704} (\bibinfo {year} {2009})}\BibitemShut {NoStop}%
\bibitem [{\citenamefont {Sofo}\ \emph {et~al.}(2012)\citenamefont {Sofo},
  \citenamefont {Usaj}, \citenamefont {Cornaglia}, \citenamefont {Suarez},
  \citenamefont {Hern\'andez-Nieves},\ and\ \citenamefont
  {Balseiro}}]{magstrhydgrap}%
  \BibitemOpen
  \bibfield  {author} {\bibinfo {author} {\bibfnamefont {J.~O.}\ \bibnamefont
  {Sofo}}, \bibinfo {author} {\bibfnamefont {G.}~\bibnamefont {Usaj}}, \bibinfo
  {author} {\bibfnamefont {P.~S.}\ \bibnamefont {Cornaglia}}, \bibinfo {author}
  {\bibfnamefont {A.~M.}\ \bibnamefont {Suarez}}, \bibinfo {author}
  {\bibfnamefont {A.~D.}\ \bibnamefont {Hern\'andez-Nieves}}, \ and\ \bibinfo
  {author} {\bibfnamefont {C.~A.}\ \bibnamefont {Balseiro}},\ }\href {\doibase
  10.1103/PhysRevB.85.115405} {\bibfield  {journal} {\bibinfo  {journal} {Phys.
  Rev. B}\ }\textbf {\bibinfo {volume} {85}},\ \bibinfo {pages} {115405}
  (\bibinfo {year} {2012})}\BibitemShut {NoStop}%
\bibitem [{\citenamefont {Moaied}\ \emph {et~al.}(2014)\citenamefont {Moaied},
  \citenamefont {Alvarez},\ and\ \citenamefont {Palacios}}]{ferrgrasurf}%
  \BibitemOpen
  \bibfield  {author} {\bibinfo {author} {\bibfnamefont {M.}~\bibnamefont
  {Moaied}}, \bibinfo {author} {\bibfnamefont {J.~V.}\ \bibnamefont {Alvarez}},
  \ and\ \bibinfo {author} {\bibfnamefont {J.~J.}\ \bibnamefont {Palacios}},\
  }\href {\doibase 10.1103/PhysRevB.90.115441} {\bibfield  {journal} {\bibinfo
  {journal} {Phys. Rev. B}\ }\textbf {\bibinfo {volume} {90}},\ \bibinfo
  {pages} {115441} (\bibinfo {year} {2014})}\BibitemShut {NoStop}%
\bibitem [{\citenamefont {Gmitra}\ \emph {et~al.}(2013)\citenamefont {Gmitra},
  \citenamefont {Kochan},\ and\ \citenamefont {Fabian}}]{SOChydgrap}%
  \BibitemOpen
  \bibfield  {author} {\bibinfo {author} {\bibfnamefont {M.}~\bibnamefont
  {Gmitra}}, \bibinfo {author} {\bibfnamefont {D.}~\bibnamefont {Kochan}}, \
  and\ \bibinfo {author} {\bibfnamefont {J.}~\bibnamefont {Fabian}},\ }\href
  {\doibase 10.1103/PhysRevLett.110.246602} {\bibfield  {journal} {\bibinfo
  {journal} {Phys. Rev. Lett.}\ }\textbf {\bibinfo {volume} {110}},\ \bibinfo
  {pages} {246602} (\bibinfo {year} {2013})}\BibitemShut {NoStop}%
\bibitem [{\citenamefont {Lee}\ and\ \citenamefont {Lee}(2019)}]{topohydgrap}%
  \BibitemOpen
  \bibfield  {author} {\bibinfo {author} {\bibfnamefont {K.~W.}\ \bibnamefont
  {Lee}}\ and\ \bibinfo {author} {\bibfnamefont {C.~E.}\ \bibnamefont {Lee}},\
  }\href {\doibase https://doi.org/10.1016/j.cap.2018.11.014} {\bibfield
  {journal} {\bibinfo  {journal} {Curr. Appl. Phys.}\ }\textbf {\bibinfo
  {volume} {19}},\ \bibinfo {pages} {137} (\bibinfo {year} {2019})}\BibitemShut
  {NoStop}%
\bibitem [{\citenamefont {Elias}\ \emph {et~al.}(2009)\citenamefont {Elias},
  \citenamefont {Nair}, \citenamefont {Mohiuddin}, \citenamefont {Morozov},
  \citenamefont {Blake}, \citenamefont {Halsall}, \citenamefont {Ferrari},
  \citenamefont {Boukhvalov}, \citenamefont {Katsnelson}, \citenamefont
  {Geim},\ and\ \citenamefont {Novoselov}}]{congraprohyd}%
  \BibitemOpen
  \bibfield  {author} {\bibinfo {author} {\bibfnamefont {D.~C.}\ \bibnamefont
  {Elias}}, \bibinfo {author} {\bibfnamefont {R.~R.}\ \bibnamefont {Nair}},
  \bibinfo {author} {\bibfnamefont {T.~M.~G.}\ \bibnamefont {Mohiuddin}},
  \bibinfo {author} {\bibfnamefont {S.~V.}\ \bibnamefont {Morozov}}, \bibinfo
  {author} {\bibfnamefont {P.}~\bibnamefont {Blake}}, \bibinfo {author}
  {\bibfnamefont {M.~P.}\ \bibnamefont {Halsall}}, \bibinfo {author}
  {\bibfnamefont {A.~C.}\ \bibnamefont {Ferrari}}, \bibinfo {author}
  {\bibfnamefont {D.~W.}\ \bibnamefont {Boukhvalov}}, \bibinfo {author}
  {\bibfnamefont {M.~I.}\ \bibnamefont {Katsnelson}}, \bibinfo {author}
  {\bibfnamefont {A.~K.}\ \bibnamefont {Geim}}, \ and\ \bibinfo {author}
  {\bibfnamefont {K.~S.}\ \bibnamefont {Novoselov}},\ }\href {\doibase
  10.1126/science.1167130} {\bibfield  {journal} {\bibinfo  {journal}
  {Science}\ }\textbf {\bibinfo {volume} {323}},\ \bibinfo {pages} {610}
  (\bibinfo {year} {2009})}\BibitemShut {NoStop}%
\bibitem [{\citenamefont {Balog}\ \emph {et~al.}(2010)\citenamefont {Balog},
  \citenamefont {J{\o}rgensen}, \citenamefont {Nilsson}, \citenamefont
  {Andersen}, \citenamefont {Rienks}, \citenamefont {Bianchi}, \citenamefont
  {Fanetti}, \citenamefont {L\ae{}gsgaard}, \citenamefont {Baraldi},
  \citenamefont {Lizzit}, \citenamefont {Sljivancanin}, \citenamefont
  {Besenbacher}, \citenamefont {Hammer}, \citenamefont {Pedersen},
  \citenamefont {Hofmann},\ and\ \citenamefont {Hornek\ae{}r}}]{gaphydgrap}%
  \BibitemOpen
  \bibfield  {author} {\bibinfo {author} {\bibfnamefont {R.}~\bibnamefont
  {Balog}}, \bibinfo {author} {\bibfnamefont {B.}~\bibnamefont {J{\o}rgensen}},
  \bibinfo {author} {\bibfnamefont {L.}~\bibnamefont {Nilsson}}, \bibinfo
  {author} {\bibfnamefont {M.}~\bibnamefont {Andersen}}, \bibinfo {author}
  {\bibfnamefont {E.}~\bibnamefont {Rienks}}, \bibinfo {author} {\bibfnamefont
  {M.}~\bibnamefont {Bianchi}}, \bibinfo {author} {\bibfnamefont
  {M.}~\bibnamefont {Fanetti}}, \bibinfo {author} {\bibfnamefont
  {E.}~\bibnamefont {L\ae{}gsgaard}}, \bibinfo {author} {\bibfnamefont
  {A.}~\bibnamefont {Baraldi}}, \bibinfo {author} {\bibfnamefont
  {S.}~\bibnamefont {Lizzit}}, \bibinfo {author} {\bibfnamefont
  {Z.}~\bibnamefont {Sljivancanin}}, \bibinfo {author} {\bibfnamefont
  {F.}~\bibnamefont {Besenbacher}}, \bibinfo {author} {\bibfnamefont
  {B.}~\bibnamefont {Hammer}}, \bibinfo {author} {\bibfnamefont {T.~G.}\
  \bibnamefont {Pedersen}}, \bibinfo {author} {\bibfnamefont {P.}~\bibnamefont
  {Hofmann}}, \ and\ \bibinfo {author} {\bibfnamefont {L.}~\bibnamefont
  {Hornek\ae{}r}},\ }\href {\doibase 10.1038/nmat2710} {\bibfield  {journal}
  {\bibinfo  {journal} {Nat. Mater.}\ }\textbf {\bibinfo {volume} {9}},\
  \bibinfo {pages} {315} (\bibinfo {year} {2010})}\BibitemShut {NoStop}%
\bibitem [{\citenamefont {Balakrishnan}\ \emph {et~al.}(2013)\citenamefont
  {Balakrishnan}, \citenamefont {K.~W.~Koon}, \citenamefont {Jaiswal},
  \citenamefont {Castro~Neto},\ and\ \citenamefont
  {\"{O}zyilmaz}}]{enhSOChydgrap}%
  \BibitemOpen
  \bibfield  {author} {\bibinfo {author} {\bibfnamefont {J.}~\bibnamefont
  {Balakrishnan}}, \bibinfo {author} {\bibfnamefont {G.}~\bibnamefont
  {K.~W.~Koon}}, \bibinfo {author} {\bibfnamefont {M.}~\bibnamefont {Jaiswal}},
  \bibinfo {author} {\bibfnamefont {A.~H.}\ \bibnamefont {Castro~Neto}}, \ and\
  \bibinfo {author} {\bibfnamefont {B.}~\bibnamefont {\"{O}zyilmaz}},\ }\href
  {\doibase 10.1038/nphys2576} {\bibfield  {journal} {\bibinfo  {journal} {Nat.
  Phys.}\ }\textbf {\bibinfo {volume} {9}},\ \bibinfo {pages} {284} (\bibinfo
  {year} {2013})}\BibitemShut {NoStop}%
\bibitem [{\citenamefont {Gonz{\'a}lez-Herrero}\ \emph
  {et~al.}(2016)\citenamefont {Gonz{\'a}lez-Herrero}, \citenamefont
  {G{\'o}mez-Rodr{\'\i}guez}, \citenamefont {Mallet}, \citenamefont {Moaied},
  \citenamefont {Palacios}, \citenamefont {Salgado}, \citenamefont {Ugeda},
  \citenamefont {Veuillen}, \citenamefont {Yndurain},\ and\ \citenamefont
  {Brihuega}}]{atomcontr}%
  \BibitemOpen
  \bibfield  {author} {\bibinfo {author} {\bibfnamefont {H.}~\bibnamefont
  {Gonz{\'a}lez-Herrero}}, \bibinfo {author} {\bibfnamefont {J.~M.}\
  \bibnamefont {G{\'o}mez-Rodr{\'\i}guez}}, \bibinfo {author} {\bibfnamefont
  {P.}~\bibnamefont {Mallet}}, \bibinfo {author} {\bibfnamefont
  {M.}~\bibnamefont {Moaied}}, \bibinfo {author} {\bibfnamefont {J.~J.}\
  \bibnamefont {Palacios}}, \bibinfo {author} {\bibfnamefont {C.}~\bibnamefont
  {Salgado}}, \bibinfo {author} {\bibfnamefont {M.~M.}\ \bibnamefont {Ugeda}},
  \bibinfo {author} {\bibfnamefont {J.-Y.}\ \bibnamefont {Veuillen}}, \bibinfo
  {author} {\bibfnamefont {F.}~\bibnamefont {Yndurain}}, \ and\ \bibinfo
  {author} {\bibfnamefont {I.}~\bibnamefont {Brihuega}},\ }\href {\doibase
  10.1126/science.aad8038} {\bibfield  {journal} {\bibinfo  {journal}
  {Science}\ }\textbf {\bibinfo {volume} {352}},\ \bibinfo {pages} {437}
  (\bibinfo {year} {2016})}\BibitemShut {NoStop}%
\bibitem [{\citenamefont {{Li}}\ \emph {et~al.}(2019)\citenamefont {{Li}},
  \citenamefont {{Yu}}, \citenamefont {{Gao}},\ and\ \citenamefont
  {{Xie}}}]{Li2019}%
  \BibitemOpen
  \bibfield  {author} {\bibinfo {author} {\bibfnamefont {S.}~\bibnamefont
  {{Li}}}, \bibinfo {author} {\bibfnamefont {R.}~\bibnamefont {{Yu}}}, \bibinfo
  {author} {\bibfnamefont {J.-H.}\ \bibnamefont {{Gao}}}, \ and\ \bibinfo
  {author} {\bibfnamefont {X.~C.}\ \bibnamefont {{Xie}}},\ }\href@noop {} {\ ,\
  \bibinfo {eid} {arXiv:1911.07006} (\bibinfo {year} {2019})}\BibitemShut
  {NoStop}%
\bibitem [{\citenamefont {Zhang}\ \emph
  {et~al.}(2017{\natexlab{b}})\citenamefont {Zhang}, \citenamefont {Li},
  \citenamefont {L\"u},\ and\ \citenamefont {Gao}}]{zhangzuquan2017}%
  \BibitemOpen
  \bibfield  {author} {\bibinfo {author} {\bibfnamefont {Z.-Q.}\ \bibnamefont
  {Zhang}}, \bibinfo {author} {\bibfnamefont {S.}~\bibnamefont {Li}}, \bibinfo
  {author} {\bibfnamefont {J.-T.}\ \bibnamefont {L\"u}}, \ and\ \bibinfo
  {author} {\bibfnamefont {J.-H.}\ \bibnamefont {Gao}},\ }\href {\doibase
  10.1103/PhysRevB.96.075410} {\bibfield  {journal} {\bibinfo  {journal} {Phys.
  Rev. B}\ }\textbf {\bibinfo {volume} {96}},\ \bibinfo {pages} {075410}
  (\bibinfo {year} {2017}{\natexlab{b}})}\BibitemShut {NoStop}%
\bibitem [{\citenamefont {Robinson}\ \emph {et~al.}(2008)\citenamefont
  {Robinson}, \citenamefont {Schomerus}, \citenamefont {Oroszl\'any},\ and\
  \citenamefont {Fal'ko}}]{adsorcond}%
  \BibitemOpen
  \bibfield  {author} {\bibinfo {author} {\bibfnamefont {J.~P.}\ \bibnamefont
  {Robinson}}, \bibinfo {author} {\bibfnamefont {H.}~\bibnamefont {Schomerus}},
  \bibinfo {author} {\bibfnamefont {L.}~\bibnamefont {Oroszl\'any}}, \ and\
  \bibinfo {author} {\bibfnamefont {V.~I.}\ \bibnamefont {Fal'ko}},\ }\href
  {\doibase 10.1103/PhysRevLett.101.196803} {\bibfield  {journal} {\bibinfo
  {journal} {Phys. Rev. Lett.}\ }\textbf {\bibinfo {volume} {101}},\ \bibinfo
  {pages} {196803} (\bibinfo {year} {2008})}\BibitemShut {NoStop}%
\bibitem [{\citenamefont {Dalla~Torre}\ \emph
  {et~al.}(2016{\natexlab{b}})\citenamefont {Dalla~Torre}, \citenamefont
  {Benjamin}, \citenamefont {He}, \citenamefont {Dentelski},\ and\
  \citenamefont {Demler}}]{FOproquasi}%
  \BibitemOpen
  \bibfield  {author} {\bibinfo {author} {\bibfnamefont {E.~G.}\ \bibnamefont
  {Dalla~Torre}}, \bibinfo {author} {\bibfnamefont {D.}~\bibnamefont
  {Benjamin}}, \bibinfo {author} {\bibfnamefont {Y.}~\bibnamefont {He}},
  \bibinfo {author} {\bibfnamefont {D.}~\bibnamefont {Dentelski}}, \ and\
  \bibinfo {author} {\bibfnamefont {E.}~\bibnamefont {Demler}},\ }\href
  {\doibase 10.1103/PhysRevB.93.205117} {\bibfield  {journal} {\bibinfo
  {journal} {Phys. Rev. B}\ }\textbf {\bibinfo {volume} {93}},\ \bibinfo
  {pages} {205117} (\bibinfo {year} {2016}{\natexlab{b}})}\BibitemShut
  {NoStop}%
\bibitem [{\citenamefont {Chen}\ \emph {et~al.}(2011)\citenamefont {Chen},
  \citenamefont {Li}, \citenamefont {Cullen}, \citenamefont {Williams},\ and\
  \citenamefont {Fuhrer}}]{tunkondgrap}%
  \BibitemOpen
  \bibfield  {author} {\bibinfo {author} {\bibfnamefont {J.-H.}\ \bibnamefont
  {Chen}}, \bibinfo {author} {\bibfnamefont {L.}~\bibnamefont {Li}}, \bibinfo
  {author} {\bibfnamefont {W.~G.}\ \bibnamefont {Cullen}}, \bibinfo {author}
  {\bibfnamefont {E.~D.}\ \bibnamefont {Williams}}, \ and\ \bibinfo {author}
  {\bibfnamefont {M.~S.}\ \bibnamefont {Fuhrer}},\ }\href {\doibase
  10.1038/nphys1962} {\bibfield  {journal} {\bibinfo  {journal} {Nat. Phys.}\
  }\textbf {\bibinfo {volume} {7}},\ \bibinfo {pages} {535} (\bibinfo {year}
  {2011})}\BibitemShut {NoStop}%
\bibitem [{\citenamefont {Ren}\ \emph {et~al.}(2014)\citenamefont {Ren},
  \citenamefont {Guo}, \citenamefont {Pan}, \citenamefont {Zhang},
  \citenamefont {Wu}, \citenamefont {Luo}, \citenamefont {Du}, \citenamefont
  {Pantelides},\ and\ \citenamefont {Gao}}]{kondadagrap}%
  \BibitemOpen
  \bibfield  {author} {\bibinfo {author} {\bibfnamefont {J.}~\bibnamefont
  {Ren}}, \bibinfo {author} {\bibfnamefont {H.}~\bibnamefont {Guo}}, \bibinfo
  {author} {\bibfnamefont {J.}~\bibnamefont {Pan}}, \bibinfo {author}
  {\bibfnamefont {Y.~Y.}\ \bibnamefont {Zhang}}, \bibinfo {author}
  {\bibfnamefont {X.}~\bibnamefont {Wu}}, \bibinfo {author} {\bibfnamefont
  {H.-G.}\ \bibnamefont {Luo}}, \bibinfo {author} {\bibfnamefont
  {S.}~\bibnamefont {Du}}, \bibinfo {author} {\bibfnamefont {S.~T.}\
  \bibnamefont {Pantelides}}, \ and\ \bibinfo {author} {\bibfnamefont {H.-J.}\
  \bibnamefont {Gao}},\ }\href {\doibase 10.1021/nl501425n} {\bibfield
  {journal} {\bibinfo  {journal} {Nano Lett.}\ }\textbf {\bibinfo {volume}
  {14}},\ \bibinfo {pages} {4011} (\bibinfo {year} {2014})}\BibitemShut
  {NoStop}%
\bibitem [{\citenamefont {Zhuang}\ \emph {et~al.}(2009)\citenamefont {Zhuang},
  \citenamefont {Sun},\ and\ \citenamefont {Xie}}]{xieKondo}%
  \BibitemOpen
  \bibfield  {author} {\bibinfo {author} {\bibfnamefont {H.-B.}\ \bibnamefont
  {Zhuang}}, \bibinfo {author} {\bibfnamefont {Q.-F.}\ \bibnamefont {Sun}}, \
  and\ \bibinfo {author} {\bibfnamefont {X.~C.}\ \bibnamefont {Xie}},\ }\href
  {http://stacks.iop.org/0295-5075/86/i=5/a=58004} {\bibfield  {journal}
  {\bibinfo  {journal} {Europhys. Lett.}\ }\textbf {\bibinfo {volume} {86}},\
  \bibinfo {pages} {58004} (\bibinfo {year} {2009})}\BibitemShut {NoStop}%
\bibitem [{\citenamefont {Uchoa}\ \emph {et~al.}(2011)\citenamefont {Uchoa},
  \citenamefont {Rappoport},\ and\ \citenamefont {Castro~Neto}}]{kondmaggrap}%
  \BibitemOpen
  \bibfield  {author} {\bibinfo {author} {\bibfnamefont {B.}~\bibnamefont
  {Uchoa}}, \bibinfo {author} {\bibfnamefont {T.~G.}\ \bibnamefont
  {Rappoport}}, \ and\ \bibinfo {author} {\bibfnamefont {A.~H.}\ \bibnamefont
  {Castro~Neto}},\ }\href {\doibase 10.1103/PhysRevLett.106.016801} {\bibfield
  {journal} {\bibinfo  {journal} {Phys. Rev. Lett.}\ }\textbf {\bibinfo
  {volume} {106}},\ \bibinfo {pages} {016801} (\bibinfo {year}
  {2011})}\BibitemShut {NoStop}%
\bibitem [{\citenamefont {Jacob}\ and\ \citenamefont
  {Kotliar}(2010)}]{orbtunkondgrap}%
  \BibitemOpen
  \bibfield  {author} {\bibinfo {author} {\bibfnamefont {D.}~\bibnamefont
  {Jacob}}\ and\ \bibinfo {author} {\bibfnamefont {G.}~\bibnamefont
  {Kotliar}},\ }\href {\doibase 10.1103/PhysRevB.82.085423} {\bibfield
  {journal} {\bibinfo  {journal} {Phys. Rev. B}\ }\textbf {\bibinfo {volume}
  {82}},\ \bibinfo {pages} {085423} (\bibinfo {year} {2010})}\BibitemShut
  {NoStop}%
\bibitem [{\citenamefont {Power}\ and\ \citenamefont
  {Ferreira}(2011)}]{spagrap}%
  \BibitemOpen
  \bibfield  {author} {\bibinfo {author} {\bibfnamefont {S.~R.}\ \bibnamefont
  {Power}}\ and\ \bibinfo {author} {\bibfnamefont {M.~S.}\ \bibnamefont
  {Ferreira}},\ }\href {\doibase 10.1103/PhysRevB.83.155432} {\bibfield
  {journal} {\bibinfo  {journal} {Phys. Rev. B}\ }\textbf {\bibinfo {volume}
  {83}},\ \bibinfo {pages} {155432} (\bibinfo {year} {2011})}\BibitemShut
  {NoStop}%
\bibitem [{\citenamefont {Pereira}\ \emph {et~al.}(2006)\citenamefont
  {Pereira}, \citenamefont {Guinea}, \citenamefont {Lopes~dos Santos},
  \citenamefont {Peres},\ and\ \citenamefont {Castro~Neto}}]{dislocgrap}%
  \BibitemOpen
  \bibfield  {author} {\bibinfo {author} {\bibfnamefont {V.~M.}\ \bibnamefont
  {Pereira}}, \bibinfo {author} {\bibfnamefont {F.}~\bibnamefont {Guinea}},
  \bibinfo {author} {\bibfnamefont {J.~M.~B.}\ \bibnamefont {Lopes~dos
  Santos}}, \bibinfo {author} {\bibfnamefont {N.~M.~R.}\ \bibnamefont {Peres}},
  \ and\ \bibinfo {author} {\bibfnamefont {A.~H.}\ \bibnamefont
  {Castro~Neto}},\ }\href {\doibase 10.1103/PhysRevLett.96.036801} {\bibfield
  {journal} {\bibinfo  {journal} {Phys. Rev. Lett.}\ }\textbf {\bibinfo
  {volume} {96}},\ \bibinfo {pages} {036801} (\bibinfo {year}
  {2006})}\BibitemShut {NoStop}%
\bibitem [{\citenamefont {Wehling}\ \emph {et~al.}(2010)\citenamefont
  {Wehling}, \citenamefont {Yuan}, \citenamefont {Lichtenstein}, \citenamefont
  {Geim},\ and\ \citenamefont {Katsnelson}}]{resscagrap}%
  \BibitemOpen
  \bibfield  {author} {\bibinfo {author} {\bibfnamefont {T.~O.}\ \bibnamefont
  {Wehling}}, \bibinfo {author} {\bibfnamefont {S.}~\bibnamefont {Yuan}},
  \bibinfo {author} {\bibfnamefont {A.~I.}\ \bibnamefont {Lichtenstein}},
  \bibinfo {author} {\bibfnamefont {A.~K.}\ \bibnamefont {Geim}}, \ and\
  \bibinfo {author} {\bibfnamefont {M.~I.}\ \bibnamefont {Katsnelson}},\ }\href
  {\doibase 10.1103/PhysRevLett.105.056802} {\bibfield  {journal} {\bibinfo
  {journal} {Phys. Rev. Lett.}\ }\textbf {\bibinfo {volume} {105}},\ \bibinfo
  {pages} {056802} (\bibinfo {year} {2010})}\BibitemShut {NoStop}%
\bibitem [{\citenamefont {Wehling}\ \emph {et~al.}(2007)\citenamefont
  {Wehling}, \citenamefont {Balatsky}, \citenamefont {Katsnelson},
  \citenamefont {Lichtenstein}, \citenamefont {Scharnberg},\ and\ \citenamefont
  {Wiesendanger}}]{locelegrap}%
  \BibitemOpen
  \bibfield  {author} {\bibinfo {author} {\bibfnamefont {T.~O.}\ \bibnamefont
  {Wehling}}, \bibinfo {author} {\bibfnamefont {A.~V.}\ \bibnamefont
  {Balatsky}}, \bibinfo {author} {\bibfnamefont {M.~I.}\ \bibnamefont
  {Katsnelson}}, \bibinfo {author} {\bibfnamefont {A.~I.}\ \bibnamefont
  {Lichtenstein}}, \bibinfo {author} {\bibfnamefont {K.}~\bibnamefont
  {Scharnberg}}, \ and\ \bibinfo {author} {\bibfnamefont {R.}~\bibnamefont
  {Wiesendanger}},\ }\href {\doibase 10.1103/PhysRevB.75.125425} {\bibfield
  {journal} {\bibinfo  {journal} {Phys. Rev. B}\ }\textbf {\bibinfo {volume}
  {75}},\ \bibinfo {pages} {125425} (\bibinfo {year} {2007})}\BibitemShut
  {NoStop}%
\bibitem [{\citenamefont {Wehling}\ \emph {et~al.}(2009)\citenamefont
  {Wehling}, \citenamefont {Katsnelson},\ and\ \citenamefont
  {Lichtenstein}}]{adsorgrap}%
  \BibitemOpen
  \bibfield  {author} {\bibinfo {author} {\bibfnamefont {T.~O.}\ \bibnamefont
  {Wehling}}, \bibinfo {author} {\bibfnamefont {M.~I.}\ \bibnamefont
  {Katsnelson}}, \ and\ \bibinfo {author} {\bibfnamefont {A.~I.}\ \bibnamefont
  {Lichtenstein}},\ }\href {\doibase
  https://doi.org/10.1016/j.cplett.2009.06.005} {\bibfield  {journal} {\bibinfo
   {journal} {Chem. Phys. Lett.}\ }\textbf {\bibinfo {volume} {476}},\ \bibinfo
  {pages} {125} (\bibinfo {year} {2009})}\BibitemShut {NoStop}%
\end{thebibliography}%
\balance

\end{document}